\documentclass{article}
\usepackage{enumitem}
\usepackage[utf8]{inputenc}
\usepackage[T1]{fontenc}
\usepackage{lineno,hyperref}
\usepackage{amssymb, epsfig,amssymb,latexsym}
\usepackage{amsfonts,psfrag,amsmath,bbm,color,url}
\usepackage[top=3cm, bottom=3cm, left=2cm, right=2cm]{geometry} 
\usepackage{epstopdf}
\usepackage{tikz,siunitx,float}
\usepackage{algorithm}
\usepackage{algorithmic}
\usepackage{comment}
\usepackage{authblk}

\usepackage{subcaption}

\DeclareMathOperator*{\argmin}{arg\,min}

\def\PP{{{\rm l}\kern - .15em {\rm P} }}
\def\PN2{{\PP_{N}-\PP_{N-2}}}

\newcommand{\bphi}{\boldsymbol{\varphi}}

\newcommand{\deleted}[1]{{}}

\usetikzlibrary{fit,positioning}

\definecolor{vargreen}{rgb}{0.0, 0.5, 0.0}
\definecolor{orange}{RGB}{255,127,0}

\title{Hybrid ROM-PINN Framework for Closure Modeling in Convection-Dominated Systems}

\author[1]{Ferhat Kaya}
\author[2]{Birgul Koc}
\author[1]{Atakan Aygun}
\author[1]{Onur Ata}
\author[1]{Ali Karakus}

\affil[1]{\textit{\small{Department of Mechanical Engineering, Middle East Technical University, Ankara, Turkey 06800}}}
\affil[2]{\textit{\small{Departamento EDAN, Universidad de Sevilla, Seville, Spain 41004}}}

\begin{document}

\maketitle

\begin{abstract}
Reduced-order models (ROMs) have become an essential tool for reducing the computational cost of fluid flow simulations. While standard ROMs can efficiently approximate laminar flows, their accuracy often suffers in convection-dominated regimes due to the truncation of dynamically important modes. 
To account for the influence of unresolved scales, ROM closure models are commonly introduced. Classical closure strategies are typically based on phenomenological arguments or analogies with large eddy simulation (LES), often formulated within a variational multiscale (VMS) framework, in which the resolved and unresolved scales are explicitly separated and their interactions are systematically modeled. 
More recently, advances in data-driven modeling and machine learning have opened new opportunities to construct ROM closures that are both more accurate and more consistent with the underlying physics.

In this work, we develop a new ROM closure that combines machine learning with physics-based modeling principles. The closure term is derived within a VMS framework, where the reduced solution space is decomposed into resolved and unresolved components.
This VMS-derived closure term is then modeled using Physics-Informed Neural Networks (PINNs) and incorporated into a newly constructed C-PINN-ROM. The resulting closure leverages high-fidelity data while enforcing physical constraints imposed by the reduced-order equations, thereby ensuring consistency with the underlying dynamics and enhanced robustness in convection-dominated regimes.

Through this PINN-based framework, we demonstrate how physics-informed machine learning can substantially improve the accuracy and robustness of ROMs, effectively bridging classical multiscale closure modeling with state-of-the-art data-driven methodologies.
\end{abstract}


\section{Introduction} \label{sec:intro}
Accurate numerical simulation of fluid flows often requires solving the governing equations using full-order models (FOMs) with millions or even billions of degrees of freedom. Although such high-fidelity simulations can deliver highly accurate results, they remain computationally demanding. Reduced-order models (ROMs) have emerged as an efficient alternative, enabling the analysis of complex flow phenomena with significantly reduced computational cost.

A key advantage of modern ROM techniques is their offline-online decomposition: the most computationally demanding tasks are performed during an offline stage. In contrast, the online stage consists of low-dimensional and inexpensive computations for evaluating new flow states. Owing to this efficiency, ROMs are particularly attractive for applications requiring repeated simulations, such as design optimization, uncertainty quantification, optimal control, inverse problems, and data assimilation \cite{gunzburger2017ensemble,hesthaven2015certified,HLB96,noack2011reduced,quarteroni2015reduced}.

Without loss of generality, consider a nonlinear dynamical system
\begin{align}
\dot{\boldsymbol u} = \boldsymbol f(\boldsymbol u),
\label{eqn:strong_formulation}
\end{align}
with the corresponding weak formulation
\begin{align}
( \dot{\boldsymbol u}, \boldsymbol v ) = ( \boldsymbol f(\boldsymbol u), \boldsymbol v ),
\quad \forall \boldsymbol v \in \boldsymbol X,
\label{eqn:weak_formulation}
\end{align}
where $\boldsymbol f$ is a nonlinear operator and $\boldsymbol X$ is a suitable infinite-dimensional function space.

During the offline stage, the FOM is solved for selected parameter values to generate a reduced basis
$\{\boldsymbol{\varphi}_1,\ldots,\boldsymbol{\varphi}_r\}$.
Projecting the governing equations onto this low-dimensional subspace leads to the compact Galerkin reduced-order model (G-ROM) dynamical system
\begin{align}
\dot{\boldsymbol a}_r = \boldsymbol F(\boldsymbol a_r),
\label{eqn:compact_grom}
\end{align}
where $\boldsymbol a_r$ denotes the reduced coefficients and $\boldsymbol F$ represents the preassembled reduced operators. 
The resulting online simulations typically require several orders of magnitude fewer computational cost than their full-order counterparts.

Despite their efficiency, standard Galerkin ROM \eqref{eqn:compact_grom} often exhibit poor performance in convection-dominated regimes, such as turbulent or transitional flows. These systems require a large number of modes to accurately represent the dynamics, while practical ROMs retain only a small subset to preserve efficiency. This truncation often leads to spurious oscillations and a significant loss of accuracy.

To address these limitations, two main strategies have been developed: numerical stabilization techniques and ROM closure modeling.

Stabilization approaches include projection-based stabilization
\cite{AzaiezJCP21,ChaconCMAME22,NovoRubinoSINUM21},
subspace rotation \cite{balajewicz2016minimal},
variational multiscale (VMS) methods
\cite{bergmann2009enablers,iliescu2013variational,iliescu2014variational,reyes2020projection},
SUPG stabilization \cite{parish2020adjoint},
filtering techniques \cite{girfoglio2021pod},
and the enforcement of physical constraints \cite{sanderse2020non}.

An alternative and complementary approach is the ROM closure modeling, in which additional terms are introduced to capture the effect of discarded modes
$\{\boldsymbol{\varphi}_{r+1},\ldots,\boldsymbol{\varphi}_R\}$ on the resolved dynamics:
\begin{align}
\dot{\boldsymbol a}_r  
= \boldsymbol F(\boldsymbol a_r) + \mathbb{C}(\boldsymbol a_r),
\label{eqn:rom_closure}
\end{align}
where $\mathbb{C}(\boldsymbol a_r)$ represents the \emph{ROM closure (or correction) term}, explicitly accounting for the influence of unresolved modes on the reduced system.

The ROM closure problem is closely related to subgrid-scale modeling in large eddy simulation (LES). Although LES closures benefit from a well-established theoretical foundation based on turbulence theory \cite{sagaut2006large}, ROM closure modeling lacks a comparable physical framework, since ROM modes are problem-dependent and not associated with a spectral separation of scales.

The VMS framework \cite{Hughes98} provides a systematic approach to closure modeling by decomposing the solution into resolved and unresolved components. VMS-ROMs naturally give rise to a closure term that must be approximated using only reduced-order information.

Recent years have witnessed a paradigm shift toward data-driven ROM closures. In \cite{koc2025residual,mou2021data,xie2018data}, classical phenomenological closures were replaced by data-driven models inferred directly from high-fidelity data, leading to the data-driven VMS-ROM. These approaches postulate parametric forms for the closure term and identify their parameters via regression. Subsequent developments included physical constraints \cite{mohebujjaman2019physically}, theoretical guaranties \cite{koc2022verifiability}, and probabilistic inference frameworks \cite{mou2023efficient}.

More recently, machine learning techniques, including Physics-Informed Neural Networks (PINNs), have been used to obtain the solution for various forward and inverse problems by incorporating physical laws into neural networks \cite{raissi_pinn_2019}. They have been used for a wide range of applications \cite{karniadakis_physics-informed_2021} and have numerous extensions \cite{jagtap_extended_2020,wang_respecting_2024, wang_piratenets_2024}. They have also been proposed to construct ROM closures that combine data-driven expressiveness with embedded physical structure \cite{ahmed2023physics,xie2020closure}. These approaches form the foundation of the present work.

In this study, by leveraging the VMS-ROM framework, the closure model is approximated utilizing a PINN architecture. The constructed network will be optimized through the enforcement of the ODE system that models the dynamical system and the known true data form of the VMS-ROM closure model. By this way, the closure model is enforced to obey the ODE system and respect the effect of the resolved state instead of solely fitting the data.   

The remainder of this paper is organized as follows.
Section~\ref{sec:rom} reviews the standard Galerkin reduced-order model for the incompressible Navier-Stokes equations.
Section~\ref{sec:vms_rom} introduces the variational multiscale decomposition underlying VMS-ROMs.
Section~\ref{sec:pinn_closure} presents the proposed PINN-based ROM closure framework.
Section~\ref{sec:numerical_results} reports numerical results for two benchmark problems:
(i) the two-dimensional viscous Burgers equation and
(ii) the two-dimensional flow past a cylinder.
Comparisons with the standard Galerkin ROM and projection errors are provided using high-fidelity FOM data.
Finally, Section~\ref{sec:conclusions_outlook} summarizes the main findings and outlines directions for future research.

\section{Reduced-Order Modeling} \label{sec:rom}
In this section, we present the ROM framework used in this work.
We begin with a brief overview of the governing equations for incompressible flow in Section~\ref{sec:governing_equations}.
Section~\ref{sec:Galerkin ROM} describes the construction of the Galerkin ROM, including the development of a POD-based basis and the derivation of the resulting reduced dynamical system.
Finally, Section~\ref{sec:limit_Galerkin ROM} discusses the inherent limitations of standard G-ROMs due to modal truncation and associated modeling errors, motivating the need for ROM closure strategies.

\subsection{Governing Equations} \label{sec:governing_equations}
Let $\Omega \subset \mathbb{R}^\textit{d}$ $(\textit{d}=2,3)$ be a bounded domain. 
The incompressible Navier-Stokes equations governing the evolution of the velocity field $\boldsymbol{u}(\boldsymbol{x},t)$ and pressure field $p(\boldsymbol{x},t)$ are given by
\begin{align}
\frac{\partial \boldsymbol{u}}{\partial t}
- Re^{-1} \Delta \boldsymbol{u}
+ \boldsymbol{u} \cdot \nabla \boldsymbol{u}
+ \nabla p &= \boldsymbol{0}, \label{eqn:nse_momentum} \\
\nabla \cdot \boldsymbol{u} &= 0, \label{eqn:nse_incompressibility}
\end{align}
supplemented with appropriate initial and boundary conditions. Here, $Re$ denotes the Reynolds number.

Throughout this work, we assume that the reduced velocity space is discretely divergence-free. Under this assumption, the incompressibility constraint is satisfied exactly at the reduced level, and the pressure variable does not appear explicitly in the reduced-order formulation.

Let $\boldsymbol X$ denote the Hilbert space of divergence-free velocity fields endowed with the $L^2(\Omega)$ inner product $(\cdot,\cdot)$. The weak form of \eqref{eqn:nse_momentum}-\eqref{eqn:nse_incompressibility} reads: find $\boldsymbol u(t) \in \boldsymbol X$ such that
\begin{align}
\left( \dot{\boldsymbol u}, \boldsymbol v \right)
= \left( \boldsymbol f(\boldsymbol u), \boldsymbol v \right),
\quad \forall \boldsymbol v \in \boldsymbol X,
\label{eqn:weak_nse}
\end{align}
where $\boldsymbol f(\boldsymbol u) := - Re^{-1} \Delta \boldsymbol u - \boldsymbol u \cdot \nabla \boldsymbol u$.

\subsection{Galerkin Reduced-Order Model (G-ROM)} \label{sec:Galerkin ROM}
The construction of the Galerkin reduced-order model relies on high-fidelity velocity data, obtained either from numerical simulations or experimental measurements of the incompressible Navier-Stokes equations. These data are assumed to belong to a finite element velocity space $V_h \subset [H_0^1(\Omega)]^\textit{d},$ which is discretely divergence-free.

Let
\begin{align}
Y =
\big[
\boldsymbol u^{Re_1}(t_1), \ldots, \boldsymbol u^{Re_1}(t_{M_1}),
\ldots,
\boldsymbol u^{Re_p}(t_1), \ldots, \boldsymbol u^{Re_p}(t_{M_p})
\big]
\in \mathbb{R}^{N \times \mathcal{M}},
\end{align}
denote the snapshot matrix collecting $\mathcal{M} := \sum_{k=1}^{p} M_k$ velocity snapshots obtained from the full-order simulations.

The reduced basis is constructed using Proper Orthogonal Decomposition (POD), which is optimal with respect to the $L^2(\Omega)$ inner product
\begin{align}
(\boldsymbol u,\boldsymbol v) := \int_\Omega \boldsymbol u \cdot \boldsymbol v \, dx.
\end{align}

The POD modes $\{\boldsymbol \varphi_i\}_{i=1}^{\mathcal{M}} \subset V_h$ are defined as the eigenfunctions of the snapshot correlation operator
\begin{align}
\sum_{j=1}^{\mathcal{M}} (\boldsymbol u(t_j), \boldsymbol \varphi_i)\, \boldsymbol u(t_j)
= \sigma_i^2 \boldsymbol \varphi_i,
\end{align}

and satisfy the orthogonality condition
\begin{align}
(\boldsymbol \varphi_i,\boldsymbol \varphi_j) = \delta_{ij}.
\end{align}

In practice, the POD basis is computed using a mass-weighted singular value decomposition of the snapshot matrix. This formulation is equivalent to the standard snapshot correlation eigenvalue problem.

For a prescribed ROM dimension $r \ll \mathcal{M}$, we define the reduced space
\begin{align}
\boldsymbol X_r := \text{span}\{\bphi_1,\dots,\bphi_r\} \subset V_h.
\end{align}
Let $\Pi_r : V_h \to \boldsymbol X_r$ denote the $L^2$-orthogonal projector, defined by
\begin{align}
(\boldsymbol u - \Pi_r \boldsymbol u, \boldsymbol v_r) = 0,
\quad \forall \boldsymbol v_r \in \boldsymbol X_r.
\label{eqn:l2_proj}
\end{align}

The POD projection error quantifies how well the full-order solution is represented in the reduced space:
\begin{align}
\|\boldsymbol u - \Pi_r \boldsymbol u\|_{L^2}^2
= \sum_{k=r+1}^{\mathcal{M}} \sigma_k^2,
\label{eqn:pod_error}
\end{align}
where $\sigma_k^2$ are the eigenvalues corresponding to the discarded modes. This error provides an a priori estimate of the discrepancy between the full-order solution and its ROM approximation. 
The ROM error is fully controlled by the number of retained modes $r$, and a slow decay of the discarded eigenvalues indicates that a low-dimensional ROM may not capture all significant dynamics.

The velocity field is approximated by
\begin{align}
\boldsymbol u(t) \approx \boldsymbol u_r(t) = \Pi_r \boldsymbol u(t) = \sum_{i=1}^{r} (a_r)_i(t)\, \boldsymbol \varphi_i,
\label{eqn:rom_expansion}
\end{align}
with reduced coefficients
\begin{align}
(a_r)_i(t) = (\boldsymbol u(t), \boldsymbol \varphi_i), \quad i=1,\dots,r.
\end{align}

To derive the reduced dynamics, we consider the semi-discrete form of the Navier-Stokes equations
\begin{align}
\frac{d\boldsymbol u(t)}{dt} = \boldsymbol f(\boldsymbol u(t)),
\label{eqn:fom_abstract}
\end{align}
where $\boldsymbol f : V_h \to V_h'$ denotes the discrete Navier-Stokes operator. Applying the projector $\Pi_r$ yields the exact projected dynamics
\begin{align}
(\dot{\boldsymbol u}_r, \boldsymbol v_r) = (\boldsymbol f(\boldsymbol u), \boldsymbol v_r), \quad \forall \boldsymbol v_r \in \boldsymbol X_r,
\label{eqn:exact_proj}
\end{align}
which is not closed since it depends on the full solution $\boldsymbol u$.

The Galerkin ROM is obtained by replacing $\boldsymbol u$ with its reduced approximation $\boldsymbol u_r$ in \eqref{eqn:exact_proj}, leading to
\begin{align}
(\dot{\boldsymbol u}_r, \boldsymbol v_r) = (\boldsymbol f(\boldsymbol u_r), \boldsymbol v_r), \quad \forall \boldsymbol v_r \in \boldsymbol X_r,
\label{eqn:Galerkin ROM_weak}
\end{align}
or, in modal form,
\begin{align}
(\dot{a}_r)_i(t) = \big(\boldsymbol f\big(\sum_{j=1}^{r} (a_r)_j(t) \boldsymbol \varphi_j\big), \boldsymbol \varphi_i\big), \quad i=1,\dots,r.
\label{eqn:Galerkin ROM_modal}
\end{align}

For the incompressible Navier-Stokes equations, the Galerkin ROM dynamical system can be explicitly written as
\begin{align}
\dot{\boldsymbol a}_r = A \, \boldsymbol a_r + \boldsymbol a_r^\top B \, \boldsymbol a_r, 
\label{eqn:nse_grom}
\end{align}
where $A \in \mathbb{R}^{r \times r}$ is the linear operator associated with viscous diffusion and
$B \in \mathbb{R}^{r \times r \times r}$ is a third-order tensor representing the convective nonlinearity.
The Galerkin ROM operators are defined componentwise as
\begin{equation}
\begin{aligned}
A_{im} &= -Re^{-1} (\nabla \boldsymbol \varphi_m, \nabla \boldsymbol \varphi_i), \\
B_{imn} &= -(\boldsymbol \varphi_m \cdot \nabla \boldsymbol \varphi_n, \boldsymbol \varphi_i). \label{eqn:grom_operators}
\end{aligned}
\end{equation}
This formulations \eqref{eqn:nse_grom} and \eqref{eqn:grom_operators} are compactly represented by \eqref{eqn:compact_grom}.

Algorithm~\ref{alg:Galerkin ROM} summarizes the construction of the Galerkin ROM. 
Because the POD modes are discretely divergence-free, the incompressibility constraint is satisfied exactly at the reduced level, and the pressure variable does not appear explicitly in the reduced-order formulation. 
The ROM approximation error is determined by the POD projection error \eqref{eqn:pod_error}, which depends on the eigenvalues corresponding to the discarded modes.

\begin{algorithm}[H]
\caption{Galerkin Reduced-Order Model (G-ROM)}
\label{alg:Galerkin ROM}
\begin{algorithmic}[1]
\STATE \textbf{Require:} High-fidelity snapshots $\{\boldsymbol u(t_j)\}_{j=1}^{\mathcal{M}}$, ROM dimension $r$, testing Reynolds numbers $\{Re^{\text{test}}_k\}_{k=1}^p$, final time $T$, time step $\Delta t$
\STATE \textbf{Output:} Reduced coefficients $\boldsymbol a_r(t)$ for all testing parameters
\STATE Construct snapshot matrix $Y = [\boldsymbol u(t_1),\dots,\boldsymbol u(t_{\mathcal{M}})]$
\STATE Compute POD modes $\{\boldsymbol \varphi_i\}_{i=1}^{r}$ via mass-weighted SVD
\STATE Assemble reduced operators:
\STATE \quad $A_{im} = -Re^{-1} (\nabla \boldsymbol \varphi_m, \nabla \boldsymbol \varphi_i)$
\STATE \quad $B_{imn} = -(\boldsymbol \varphi_m \cdot \nabla \boldsymbol \varphi_n, \boldsymbol \varphi_i)$
\STATE Project initial condition: $(\boldsymbol a_r)_i(0) = (\boldsymbol u(0), \boldsymbol \varphi_i), \quad i=1,\dots,r$ 
\FOR {$k = 1, \dots, p$} 
\STATE Initialize $\boldsymbol a \gets \boldsymbol a_r(0)$, $t \gets 0$
\WHILE {$t < T$}  
\STATE Update reduced coefficients:
        \STATE \quad $\boldsymbol a_r \gets \boldsymbol a_r + \Delta t \left( A \, \boldsymbol a_r + \boldsymbol a_r^\top B \, \boldsymbol a_r \right)$
        \STATE $t \gets t + \Delta t$
    \ENDWHILE
    \STATE Save $\boldsymbol a_r(t)$ for $Re^{\text{test}}_k$
\ENDFOR
\STATE \RETURN $\boldsymbol a_r(t)$
\end{algorithmic}
\end{algorithm}

\subsection{Limitations of Galerkin ROMs}
\label{sec:limit_Galerkin ROM}
Although Galerkin ROMs inherit important structural properties of the Navier-Stokes equations through projection, they suffer from intrinsic limitations due to modal truncation. In particular, the Galerkin approximation neglects the interaction between the resolved modes $\boldsymbol u_r = \Pi_r \boldsymbol u$ and the discarded modes $\boldsymbol u' = (I-\Pi_r)\boldsymbol u$, which appears as a modeling error in the reduced dynamics:
\begin{align}
(\boldsymbol f(\boldsymbol u_r + (I-\Pi_r)\boldsymbol u) - \boldsymbol f(\boldsymbol u_r), \boldsymbol v_r),
\quad \forall \boldsymbol v_r \in \boldsymbol X_r.
\end{align}

For nonlinear systems, this neglected interaction can accumulate over time and severely affect the accuracy and stability of the reduced-order solution. This issue is especially pronounced in convection-dominated and turbulent flows, where energy transfer across scales plays a central role. As a result, standard Galerkin ROMs may exhibit spurious oscillations, loss of accuracy, or even numerical instability \cite{HLB96,noack2011reduced,osth2014need,wang2012proper}.

A practical indicator of under-resolution is the decay rate of the POD eigenvalues: a slow decay implies strong multiscale interactions, suggesting that a low-dimensional linear subspace is insufficient to represent the dynamics accurately. While increasing the number of retained modes can mitigate this issue, it often undermines the computational efficiency that motivates reduced-order modeling. These observations motivate the development of systematic approaches that retain a low-dimensional resolved space while accounting for the influence of truncated modes, such as the variational multiscale framework.

\section{Variational Multiscale Reduced-Order Modeling}
\label{sec:vms_rom}
The variational multiscale framework provides a principled mechanism for identifying and correcting the deficiencies of Galerkin ROMs in under-resolved regimes. Rather than attributing instability and loss of accuracy solely to modal truncation, VMS explicitly characterizes the missing interactions between resolved and unresolved scales as the fundamental source of modeling error.

Starting from the abstract weak formulation, the VMS approach introduces an orthogonal decomposition of the solution space using the reduced-order projector $\Pi_r$ and its complement $I-\Pi_r$. The solution is decomposed as
\begin{align}
\boldsymbol u = \boldsymbol u_r + \boldsymbol u', \qquad
\boldsymbol u_r = \Pi_r \boldsymbol u \in \boldsymbol X_r, \qquad
\boldsymbol u' = (I-\Pi_r)\boldsymbol u \in \boldsymbol X_r^\perp .
\label{eq:vms_decomp}
\end{align}

Substituting~\eqref{eq:vms_decomp} into the governing equations and projecting onto the resolved space $\boldsymbol X_r$ yields
\begin{align}
\left( \dot{\boldsymbol u}_r, \boldsymbol v_r \right)
=
\left( \boldsymbol f(\boldsymbol u_r), \boldsymbol v_r \right)
+
\left( \boldsymbol f(\boldsymbol u_r + \boldsymbol u') - \boldsymbol f(\boldsymbol u_r), \boldsymbol v_r \right),
\quad
\forall \boldsymbol v_r \in \boldsymbol X_r .
\label{eq:vms_rom_weak}
\end{align}

The second term on the right-hand side of~\eqref{eq:vms_rom_weak} represents the \emph{VMS-ROM closure stress}, i.e., the net influence of unresolved scales on the resolved dynamics. This term is exact but not available during online simulations, since it depends explicitly on the unresolved component $\boldsymbol u' \in \boldsymbol X_r^\perp$.

Projecting~\eqref{eq:vms_rom_weak} onto the reduced basis $\{\boldsymbol\varphi_i\}_{i=1}^r$ yields the following dynamical system
\begin{align}
\dot{\boldsymbol a}_r
=
\boldsymbol F(\boldsymbol a_r)
+
\boldsymbol \tau_r ,
\label{eqn:vms_rom_exact}
\end{align}
where $\boldsymbol F(\boldsymbol a_r)$ denotes the standard Galerkin ROM operator, and the exact VMS-ROM closure stress vector $\boldsymbol \tau_r \in \mathbb{R}^r$ is defined componentwise as
\begin{align}
(\boldsymbol \tau_r)_i
:=
\left(
\boldsymbol f(\boldsymbol u_r + \boldsymbol u')
-
\boldsymbol f(\boldsymbol u_r),
\boldsymbol \varphi_i
\right),
\qquad i = 1,\dots,r .
\label{eq:vms_tau_def}
\end{align}

Equation~\eqref{eqn:vms_rom_exact} defines the \emph{VMS-ROM closure problem}: the reduced dynamics are exact but unclosed due to the presence of the unresolved-scale stress $\boldsymbol \tau_r$. Whenever $\boldsymbol \tau_r \neq \boldsymbol 0$, the Galerkin ROM fails to account for essential multiscale interactions induced by modal truncation.

\subsection{VMS-ROM Closure Modeling}
\label{sec:rom_closure}
Within the VMS framework, the closure problem is rigorously defined as the approximation of the VMS-ROM stress term $\boldsymbol \tau_r$ in~\eqref{eqn:vms_rom_exact}. Since $\boldsymbol u'$ is unavailable during the online stage, $\boldsymbol \tau_r$ cannot be evaluated exactly and must be modeled as a function of the resolved trajectory.

We therefore introduce the closure ansatz
\begin{align}
\boldsymbol \tau_r \;\approx\; \mathbb{C}(\boldsymbol a_r),
\label{eq:closure_ansatz}
\end{align}
where $\mathbb{C} : \mathbb{R}^r \rightarrow \mathbb{R}^r$ denotes a reduced-order closure operator.

Substituting~\eqref{eq:closure_ansatz} into~\eqref{eqn:vms_rom_exact} yields the closed reduced-order model which coincides with the general closure formulation introduced in~\eqref{eqn:rom_closure}, now grounded in a variational multiscale interpretation.

From the VMS perspective, the ROM closure term represents a modeled approximation of the projected unresolved-scale residual,
\begin{align}
\boldsymbol \tau_r
=
\Pi_r \Big(
\boldsymbol f(\boldsymbol u_r + \boldsymbol u')
-
\boldsymbol f(\boldsymbol u_r)
\Big),
\end{align}
and can be expressed through a modeling operator $g(\cdot)$ as
\begin{align}
\mathbb{C}(\boldsymbol a_r)
:=
g\!\left(
\Pi_r \big(
\boldsymbol f(\boldsymbol u_r + \boldsymbol u')
-
\boldsymbol f(\boldsymbol u_r)
\big)
\right).
\label{eq:closure_g}
\end{align}

When the operator $g(\cdot)$ depends only on resolved-scale quantities, the ROM closure model $\mathbb{C}(\boldsymbol a_r)$ becomes a function of the resolved modal trajectory alone. In this way, the ROM closure model acts as a correction to the Galerkin ROM, enabling the reduced system to account for the influence of truncated dynamics without increasing the ROM dimension.

Within this framework, different ROM closure strategies correspond to different realizations of the ROM closure model operator $\mathbb{C}(\cdot)$. Classical approaches include eddy-viscosity models and phenomenological ansatzes inspired by turbulence modeling, while modern approaches infer $\mathbb{C}$ directly from data using machine learning techniques. Importantly, the VMS formulation clarifies that $\mathbb{C}(\boldsymbol a_r)$ represents a reduced-order approximation of the unresolved-scale stress and should therefore preserve stability, consistency, and the underlying physical structure of the system.

In this work, the ROM closure model $\mathbb{C}(\boldsymbol a_r)$ is learned using a physics-informed neural networks (PINNs). By embedding the reduced-order governing equations into the learning process, the proposed approach ensures that the learned ROM closure model remains consistent with the VMS-ROM closure term while accurately capturing the influence of truncated dynamics without increasing the ROM dimension.

\section{Physics-Informed Neural Networks for ROM Closure}
\label{sec:pinn_closure}
Machine learning methods have been widely utilized in the reduced-order model studies to generate data-driven non-intrusive models and in the stabilization of the intrusive Galerkin ROMs \cite{ahmed2021closures}. While these data-driven approaches are generally easier to implement than their intrusive counterparts, they are often data-hungry and tend to perform poorly when extrapolating beyond the training regime.  In this context, PINNs have emerged as an alternative paradigm in the ROM community, as they combine the strengths of both intrusive and non-intrusive modeling approaches. With these models, the use of physical constraints reduces the cost of high-fidelity data offline. The main objective of this type of models is 
to predict the time-dependent reduced-order coefficients $\boldsymbol a_r(t;\mu)$ for a given set of parameters $\mu$ by minimizing the loss that might depend either on ROM or FOM equations coupled with data \cite{chen2021physics} \cite{fu2023physicsdatacombined}. 

Another branch of the reduced order model studies for which the machine learning architectures are widely being utilized is the stabilization of the intrusive Galerkin ROMs through closure models. In intrusive reduced order models, the VMS-ROM closure problem~\eqref{eqn:vms_rom_exact}
arises due to the truncation of the unresolved-scale modes in the system. Since there are no analytical relations that will provide the interaction between the resolved and unresolved modes, the VMS-ROM closure problem~\eqref{eqn:vms_rom_exact} is handled by appropriate modeling strategies. The reader is referred to the comprehensive review article for the different types of closure modeling strategies developed for reduced order models \cite{ahmed2021closures}.

With the aforementioned VMS formulation of the Galerkin ROMs, the interaction between the truncated and the resolved scales is explicitly obtained via the interaction between the resolved and unresolved scales. In that respect, this article benefits the PINN architecture together with the VMS formulation through the ODE loss enforcement in order to model the VMS-ROM closure term in \eqref{eqn:vms_rom_exact}. This chapter will first introduce the naive feed-forward neural network architectures and their physics-informed forms. After that, the developed PINN closure models will be explained.

\subsection{Physics Informed Neural Networks}
A basic, fully connected deep neural network architecture can be used to solve differential equations \cite{lagaris1998artificial}. Given an input vector $\mathbf{x} \in \mathbb{R}^d$, a single layer neural network gives an output $\hat{\mathbf{u}}$ by the following form:

\begin{equation}
    \label{eq:singleLayerNN}
    \hat{\mathbf{u}} = \sigma(\mathbf{W}_1\mathbf{x} + \mathbf{b}_1)\mathbf{W}_2 + \mathbf{b}_2,
\end{equation}
where $\mathbf{W}$ are the weight matrices and $\mathbf{b}$ are the bias vectors. $\sigma(\cdot)$ is a nonlinear function known as the activation function. In general, Sigmoid, hyperbolic tangent, and rectified linear unit (ReLU) are popular choices for the activation function. The hyperparameters $\theta = [\mathbf{W}, \mathbf{b}]$ are estimated by the following optimization problem

\begin{equation}
    \label{eq:optimization}
    \mathbf{\theta}^* = \argmin_\mathbf{\theta} J(\mathbf{\theta}; \mathbf{x}).
\end{equation}
Here, $J$ is the objective function to be minimized. This minimization problem in \eqref{eq:optimization} can be solved by using first-order stochastic gradient descent (SGD) algorithms. The loss function of a PINN is commonly a composite loss function composed of data and physics loss as 
\begin{align}
\mathcal{L}(\theta) = \mathcal{L}_{\text{data}} + \lambda\, \mathcal{L}_{\text{physics}},
\end{align}
where $\mathcal{L}_{\text{data}}$ is the loss of initial and/or boundary conditions as well as the labeled observations, and $\mathcal{L}_{\text{physics}}$ loss term for the residual of the differential equation. The hyperparameter $\lambda > 0$ balances the data and physics contributions which can be manually specified or tuned automatically \cite{wang_when_2022, mcclenny_self-adaptive_2023,anagnostopoulos_residual-based_2024}. The hyperparameters are optimized using commonly used ADAM algorithm \cite{kingma2014adam}.

\subsection{PINN for Galerkin ROM Closure}
The PINN closure model is constructed to enhance both the accuracy and physical consistency of data-driven VMS-ROMs by learning the influence of truncated modes on the resolved ROM dynamics. By decomposing the solution space into two complementary subspaces through the VMS formulation, as in~\eqref{eq:vms_decomp}, an additional term naturally arises in the reduced-order equations, requiring a closure contribution $\boldsymbol{\tau}_r$ to complete the resolved-scale evolution equation in~\eqref{eqn:vms_rom_exact}. Moreover, the exact form of the VMS-ROM closure term, defined in~\eqref{eq:vms_tau_def}, is available and can be computed from full-order model (FOM) projection data during the offline stage. In this context, a PINN-based closure model, $\mathbb{C}(\boldsymbol a_r)$, is developed to approximate the exact VMS-ROM closure term $\boldsymbol{\tau}_r$.

In that sense, the constructed PINN closure model predicts the time coefficients of the resolved POD modes together with the closure terms in the VMS formulation of the evolution equation to construct the ODE loss term in the model. 

\begin{figure}[htbp!]
\centering
\includegraphics[width=1\linewidth]{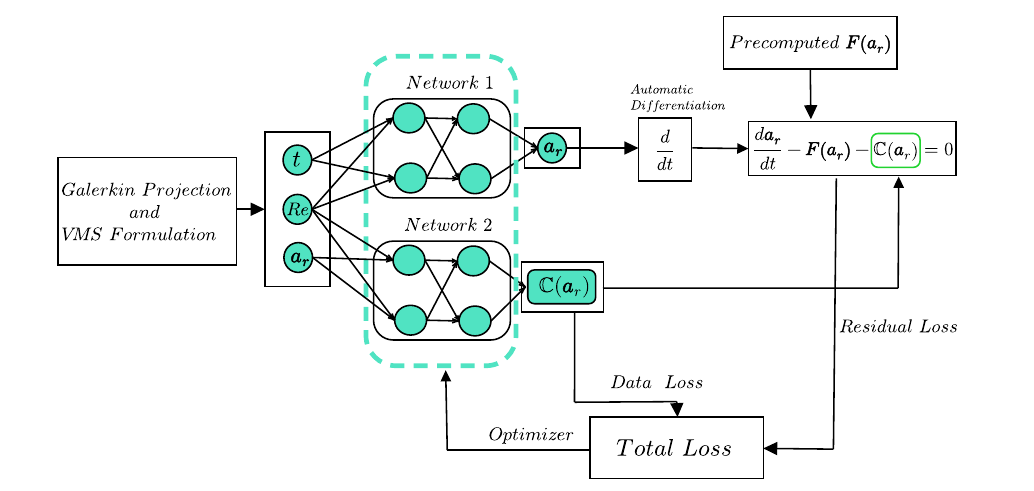}
\caption{A schematic diagram of the proposed PINN based closure model.} \label{fig:pinn_diagram}
\end{figure}

A representative diagram of the proposed closure model is shown in Figure~\ref{fig:pinn_diagram}. The Galerkin ROM augmented with this closure model is referred to as the C-PINN-ROM throughout the remainder of the paper. The inputs to the neural network consist of time, the resolved-scale modal coefficients obtained from the VMS formulation, and problem parameters such as the Reynolds number~$Re$. The network outputs the closure terms, $\mathbb{C}(\boldsymbol a_r)$, which account for both unresolved-scale effects and mixed interactions between resolved and unresolved scales in the reduced-order dynamical system.

Using the mean squared error to define both the physics and data losses, the individual terms of the composite loss function are given by:
\begin{align}
\mathcal{L}_{\text{physics}} &= \frac{1}{N_t}\sum_{i=1}^{N_t} \Big |\frac{d\boldsymbol{a}_r}{dt} - \mathbf{F}(\boldsymbol{a}_r) - \mathbb{C}(\boldsymbol a_r) \Big |^2 \label{eqn:loss_physics} \\
\mathcal{L}_{\text{data}} &= \frac{1}{N_d}\sum_{i=1}^{N_d} \Big|\boldsymbol \tau_r - \mathbb{C}(\boldsymbol a_r)  \Big|^2,  \label{eqn:loss_data}
\end{align}

In constructing the composite loss, the model leverages the known form of the predicted time coefficients and the right-hand-side terms through the data loss in~\eqref{eqn:loss_data}, while the physics-based ODE loss in~\eqref{eqn:loss_physics} enforces consistency of the network predictions with the underlying dynamical system.

\section{Numerical Results}
\label{sec:numerical_results}
In this section, the capabilities of the newly developed ROM model (C-PINN-ROM) are evaluated using two benchmark problems: the two-dimensional Burgers equation and flow past a circular cylinder. For the Burgers equation, the closure model is assessed in a parametric setting, whereas for the cylinder flow, the model’s temporal extrapolation capability is examined.

In Sections~\ref{2d_burgers_section} and~\ref{2d_fpc_section}, each benchmark problem is introduced together with the corresponding high-fidelity data collection procedures used to construct the reduced-order spaces. The primary objective of this study is to assess the accuracy of the proposed C-PINN-ROM. To this end, its performance is evaluated by comparing the predicted modal coefficients and reconstructed velocity fields with those obtained from standard Galerkin ROMs of varying subspace dimensions, as well as with the reference projection of the FOM solution onto the reduced basis. 

To quantify numerical accuracy, the average relative $L_2$ error is employed:
\begin{equation}
\mathcal{E}(L^2) = \frac{1}{M} \sum_{k=1}^{M} 
\frac{\left\| \boldsymbol{u}_{FOM}(t_k) - \boldsymbol{u}_{ROM}(t_k) \right\|_{L^2}}
{\left\| \boldsymbol{u}_{FOM}(t_k) \right\|_{L^2}},
\label{eq:rel_l2_fom_err}
\end{equation}
where the velocity fields for each ROM model are reconstructed using the corresponding modal coefficients and POD modes.

Throughout the comparisons, the proposed closure model is first evaluated against the ground-truth time coefficients and velocity fields to demonstrate the improvements over the uncorrected Galerkin ROM within the same reduced subspace. In addition, Galerkin ROMs operating in higher-dimensional modal spaces are included to illustrate the impact of increasing the subspace dimension on ROM accuracy. Finally, the PINN-ROM framework for both benchmark problems is implemented using JAX~\cite{jagtap_extended_2020} as the computational backend.

\subsection{2-Dimensional Viscous Burgers Equation}
\label{2d_burgers_section}
The newly developed C-PINN-ROM is evaluated on the coupled Burgers’ equations, a system of two partial differential equations describing the interaction between two velocity fields. This system extends the classical Burgers’ equation—an archetypal model capturing both convective and dissipative effects—by incorporating the mutual influence of the two velocity components. As a result, the coupled Burgers’ equations exhibit richer and more complex dynamics, making them an ideal testbed for the proposed closure model.

The governing equations are given in their full form as
\begin{align}
\frac{\partial u}{\partial t} + u \frac{\partial u}{\partial x} + v \frac{\partial u}{\partial y} 
&= \frac{1}{Re} \left( \frac{\partial^2 u}{\partial x^2} + \frac{\partial^2 u}{\partial y^2} \right), \\[2mm]
\frac{\partial v}{\partial t} + u \frac{\partial v}{\partial x} + v \frac{\partial v}{\partial y} 
&= \frac{1}{Re} \left( \frac{\partial^2 v}{\partial x^2} + \frac{\partial^2 v}{\partial y^2} \right),
\end{align}
with the spatial and temporal domains defined as
\begin{equation}
(x, y) \in \Omega = (0,1) \times (0,1), \quad t \in (0,1).
\end{equation}

Notably, the coupled Burgers’ equations produce a dynamical ROM system analogous to that derived from the incompressible Navier-Stokes equations in~\eqref{eqn:nse_grom}, \eqref{eqn:nse_momentum}, and \eqref{eqn:nse_incompressibility}. Consequently, the G-ROM and PINN closure model are formulated following the procedure outlined in Sections~\ref{sec:rom}-\ref{sec:vms_rom}.

For the generation of ROM data, the following exact solution of the 2D coupled Burgers’ equations is used, along with its corresponding initial and boundary conditions~\cite{sreelakshmi2024adaptive}:
\begin{align}
u(x,y,t) &= \frac{3}{4}
  - \frac{1}{4 \left[\, 1 + \exp\!\left(\frac{\mathrm{Re}}{32}(-4x + 4y - t)\right) \right]}, \\
v(x,y,t) &= \frac{3}{4}
  + \frac{1}{4 \left[\, 1 + \exp\!\left(\frac{\mathrm{Re}}{32}(-4x + 4y - t)\right) \right]} .
\end{align}

\begin{figure}[hbtp!]
\centering
\includegraphics[width=0.60\linewidth]{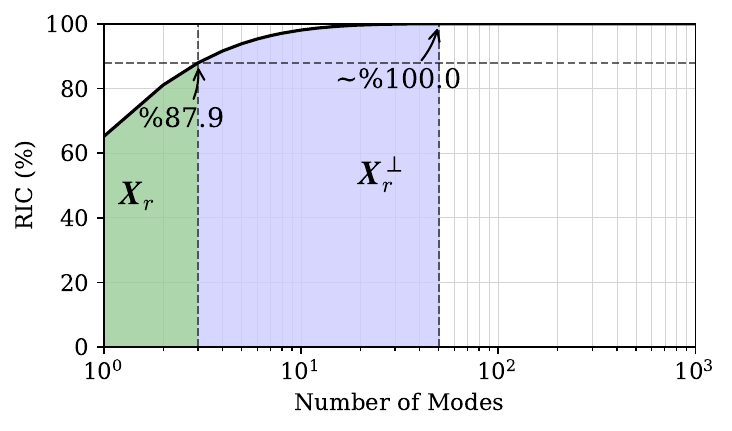}
\caption{Burgers equation; relative information content of the complementary POD subspaces $\boldsymbol{X}_r$ and $\boldsymbol{X}_r^\perp$.} \label{fig:RIC_2db}
\end{figure}

During the training of the PINN closure model, FOM data are collected for varied Reynolds numbers $Re = [1000, 2000, 3000, 4000, 5000, 6000]$, and the POD modes are computed by applying singular value decomposition (SVD) to the constructed snapshot matrix. The modal energy distribution, along with the VMS-based decomposition of the subspace into resolved and unresolved modes, is shown in Figure~\ref{fig:RIC_2db}. In this setup, the resolved subspace consists of the first three POD modes, which capture $87\%$ of the total energy. The remaining 47 modes, corresponding to the unresolved subspace and accounting for $13\%$ of the total energy, are represented through the developed closure model. Architectural details and training parameters of the PINN closure model are summarized in Table~\ref{tab:pinn_burgers_arch}.

\begin{table}[H]
\centering
\small
\begin{tabular}{l|c|c}
\hline
\textbf{Parameter} & \textbf{Network 1} & \textbf{Network 2} \\
\hline
Architecture & Fully Connected (MLP) & Fully Connected (MLP) \\
Layer Structure & $[2, 64, 64, 64, 3]$ & $[4,64,64,64,64, 15]$ \\
Activation & $\tanh$ & $\tanh$ \\
Initialization & Xavier Normal &  Xavier Normal \\
\hline
Initial Learning Rate & \multicolumn{2}{c}{$10^{-3}$} \\
Decay Rate / Steps & \multicolumn{2}{c}{$0.90$ / $5 \times 10^3$} \\
Optimizer & \multicolumn{2}{c}{Adam} \\
Training Steps & \multicolumn{2}{c}{$4\times10^5$} \\
\hline
\end{tabular}
\caption{Burgers equation; hyperparameters  and architecture of the PINN closure model $\mathbb{C}(\boldsymbol{a}_r)$.} \label{tab:pinn_burgers_arch}
\end{table}

\begin{figure}[htbp!]
\centering
\includegraphics[width=0.75\linewidth]{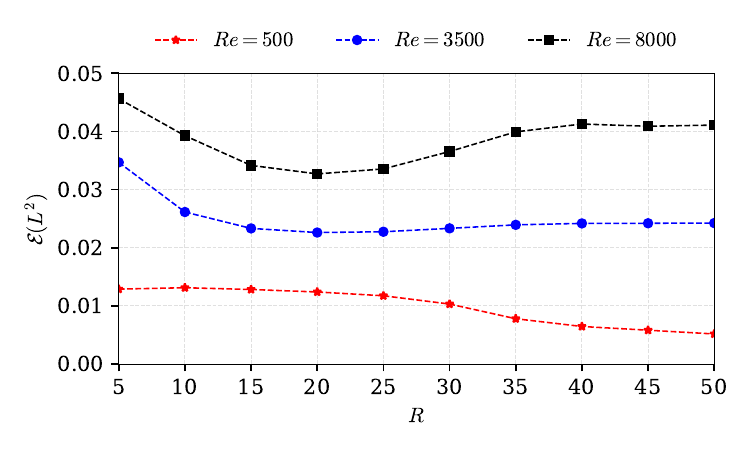}
\caption{Burgers equation; average relative $L^2$ velocity errors for different Reynolds numbers and ROM dimension.} \label{fig:L2_vs_d_2DB}
\end{figure}

As illustrated in Figure~\ref{fig:L2_vs_d_2DB}, the Galerkin ROM simulations do not exhibit a monotonic reduction in the relative $L^2$ error as the ROM subspace dimension $R$ is increased. Because the training procedure is based on projected FOM data, the same scale separation is applied consistently across all Reynolds numbers. Specifically, a fixed resolved-scale dimension of $\dim(\boldsymbol X_r)=3$ and an unresolved-scale dimension of $\dim(\boldsymbol X_r^\perp)=47$ are used throughout the study.

\begin{figure}[htbp!]
\centering
\includegraphics[width=1\linewidth]{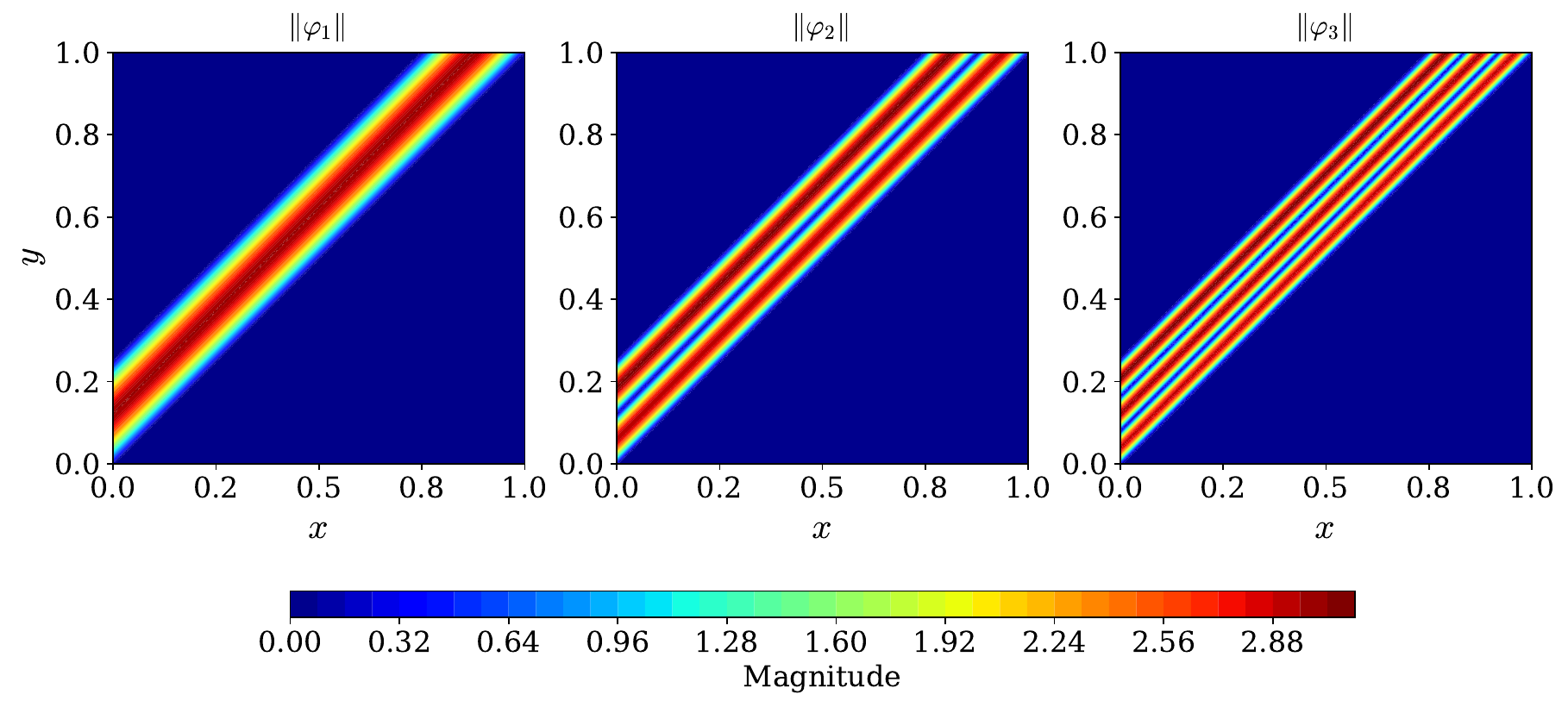}
\caption{Burgers equation; POD modes of parametric ROM subspace.} \label{fig:pod_modes_2db}
\end{figure}    

The developed PINN closure model is evaluated in both the interpolation and extrapolation regimes of the problem parameter, with interpolation Reynolds numbers $Re_{\mathrm{int}} = \{1500, 3500, 5500\}$ and extrapolation Reynolds numbers $Re_{\mathrm{ext}} = \{500, 8000\}$. The corresponding results are presented in the following subsections.

\subsubsection{Testing in Extrapolation and Interpolation Regimes}
In this section, the performance of different ROM models is evaluated based on their ability to predict the ROM time coefficients under both interpolation and extrapolation regimes. The reference time coefficients are obtained by projecting the high-fidelity solution onto the ROM subspace spanned by the three most energetic POD modes, which are shown in Figure~\ref{fig:pod_modes_2db}. These coefficients serve as the ground truth for assessing the evolution of the FOM solution within the reduced-order subspace.

In Table~\ref{tab:rom_errors_full_time_updated}, we present the average relative $L^2$ errors of G-ROMs with varying subspace dimensions ($r=3, 25, 50$), the recently developed C-PINN-ROM, and the reference projection of the full-order solution onto the reduced basis. The G-ROM with $r=3$ represents the uncorrected ROM, while the C-PINN-ROM corresponds to the ROM augmented with the developed closure model. The G-ROMs with $r=25$ and $r=50$ are included to assess the effect of increasing the subspace dimension on ROM trajectories and solution accuracy. For field reconstruction, the velocity solutions are obtained by combining the computed ROM time coefficients with the corresponding precomputed POD modes; specifically, the G-ROM with $r=3$, the C-PINN-ROM, and the reference projection all employ the first three POD modes, whereas the G-ROMs with $r=25$ and $r=50$ use the first 25 and 50 modes, respectively.

From the results, it is evident that the C-PINN-ROM significantly improves accuracy over the uncorrected G-ROM with $r=3$, reducing the relative $L^2$ error and closely approaching the ground-truth projection. While increasing the subspace dimension in G-ROMs ($r=25$ and $r=50$) also enhances accuracy, the C-PINN-ROM achieves comparable or better performance without the need for a larger reduced basis, demonstrating the effectiveness of the closure model in capturing the influence of truncated modes.

\begin{table}[htbp!]
\centering
\small
\begin{tabular}{c|c|c|c|c|c}
\hline 
\textbf{Re} & \textbf{G-ROM(r=3)} & \textbf{C-PINN-ROM(r=3)} & \textbf{Projection(r=3)} & \textbf{G-ROM(r=25)} & \textbf{G-ROM(r=50)} \\
\hline
500  & $1.34\times 10^{-2}$ & $1.25\times 10^{-2}$ & $7.61\times 10^{-3}$ & $1.17\times 10^{-2}$ & $5.18\times 10^{-3}$ \\
1500 & $2.71\times 10^{-2}$ & $1.69\times 10^{-2}$ & $1.68\times 10^{-2}$ & $2.35\times 10^{-3}$ & $2.01\times 10^{-3}$ \\
3500 & $3.78\times 10^{-2}$ & $2.15\times 10^{-2}$ & $2.15\times 10^{-2}$ & $9.76\times 10^{-3}$ & $1.23\times 10^{-2}$ \\
5500 & $4.12\times 10^{-2}$ & $2.31\times 10^{-2}$ & $2.29\times 10^{-2}$ & $1.68\times 10^{-2}$ & $2.41\times 10^{-2}$ \\
8000 & $4.32\times 10^{-2}$ & $2.40\times 10^{-2}$ & $2.37\times 10^{-2}$ & $2.36\times 10^{-2}$ & $3.60\times 10^{-2}$ \\
\hline
\end{tabular}
\caption{Burgers equation; average relative $L^2$ errors of ROM models for different Reynolds numbers}
\label{tab:rom_errors_full_time_updated}
\end{table}

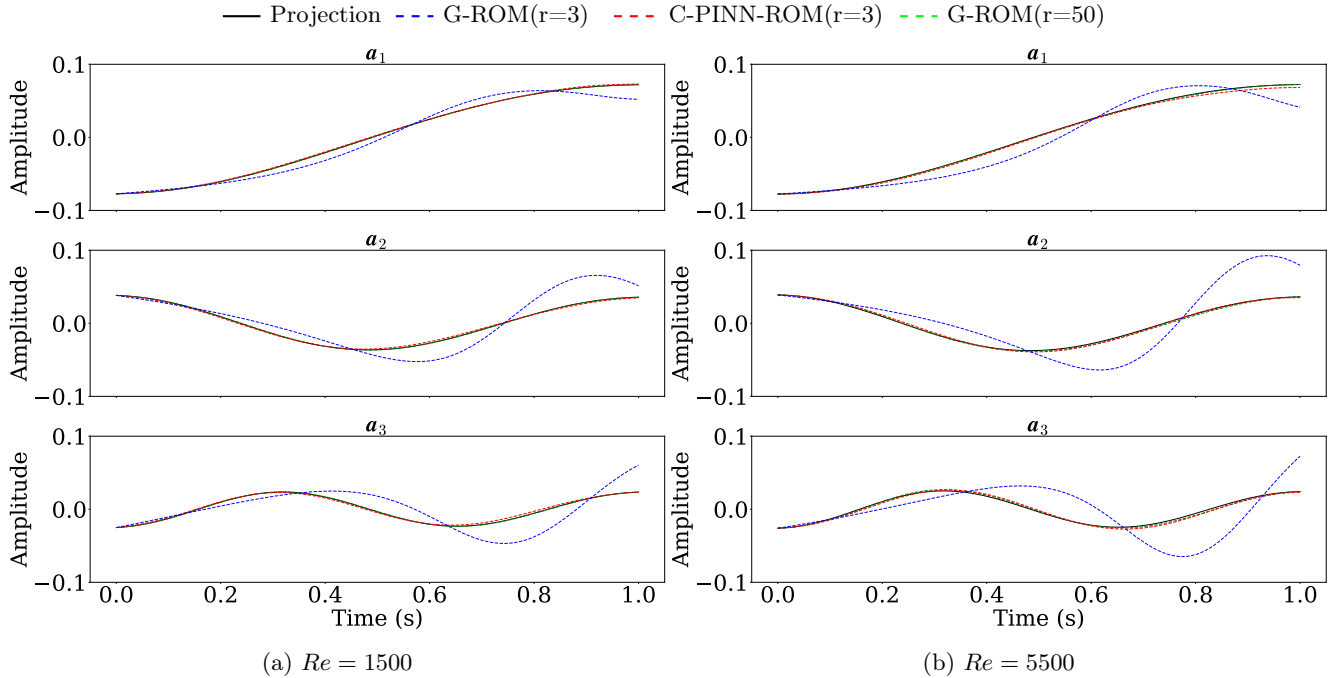
\begin{figure}[htbp!]
\centering
\begin{tikzpicture}
    \draw[thick, black] (0,0) -- (0.5,0) node[right, black] {\small Projection};
    \draw[thick, blue, dashed] (2.3,0) -- (2.8,0) node[right, black] {\small G-ROM(r=3)};
    \draw[thick, red, dashed] (5.2,0) -- (5.8,0) node[right, black] {\small C-PINN-ROM(r=3)};
    \draw[thick, green, dashed] (9.0,0) -- (9.5,0) node[right, black] {\small G-ROM(r=50)};
\end{tikzpicture}
\begin{subfigure}{.50\textwidth}
\centering
\includegraphics[width=1\linewidth]{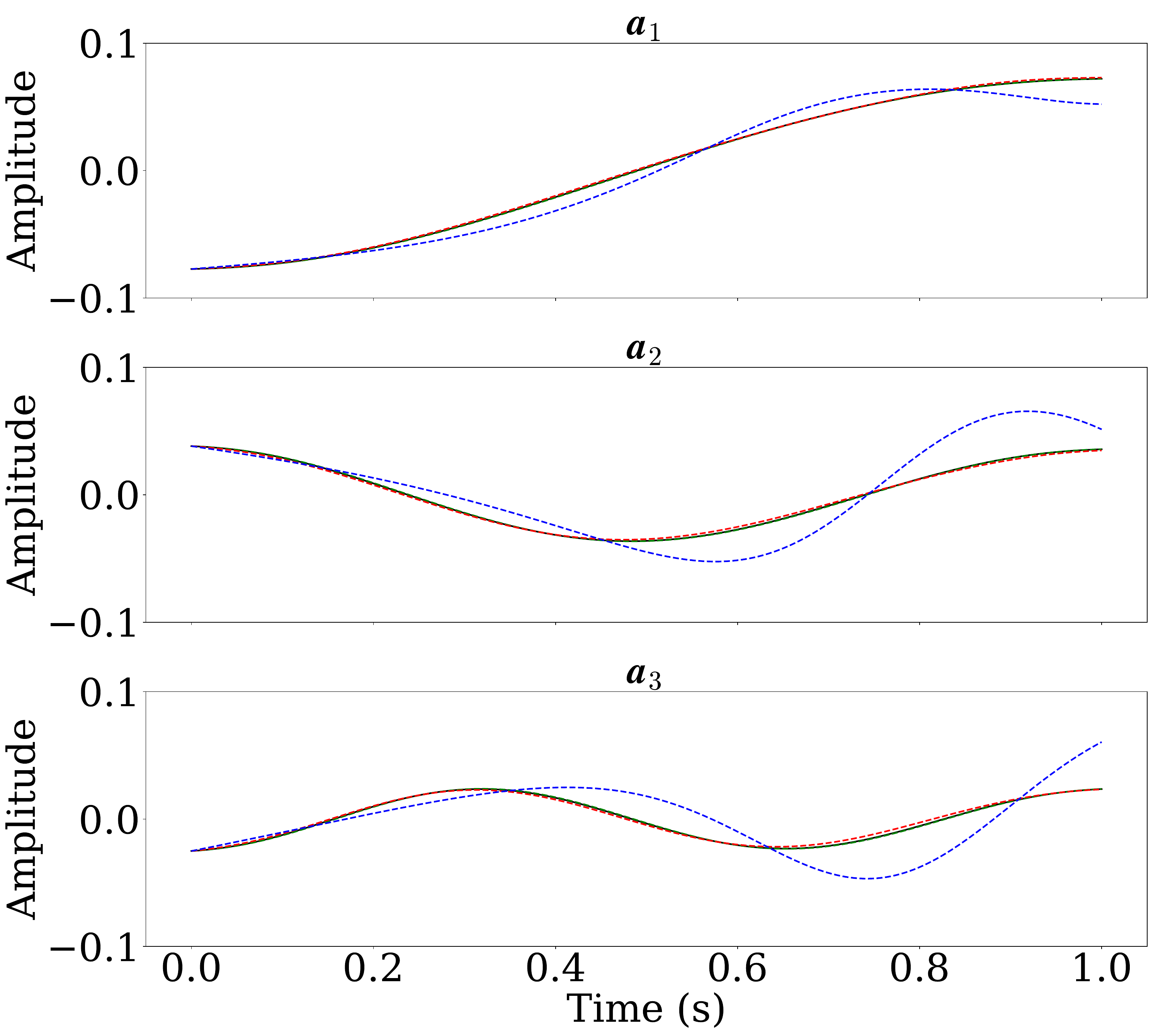}
\caption{$Re = 1500$} \label{fig:10:1500}
\end{subfigure}%
\begin{subfigure}{.50\textwidth}
\centering
\includegraphics[width=1\linewidth]{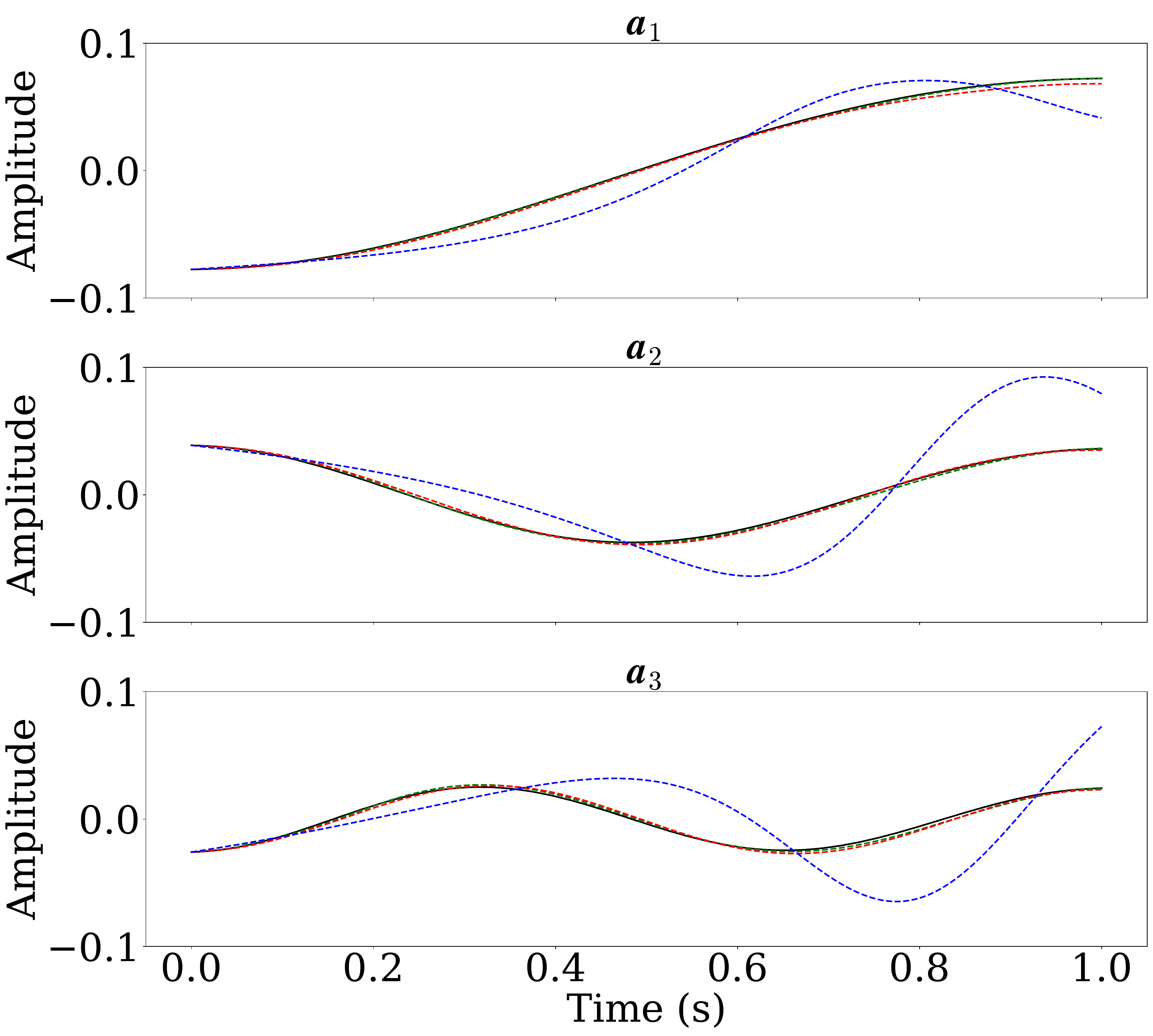}
\caption{$Re = 5500$} \label{fig:10:5500}
\end{subfigure}
\caption{Burgers equation; time-dependent coefficients in the interpolation testing regime with $Re = 1500$ and $Re=5500$.} \label{fig:Interpolation}
\end{figure}

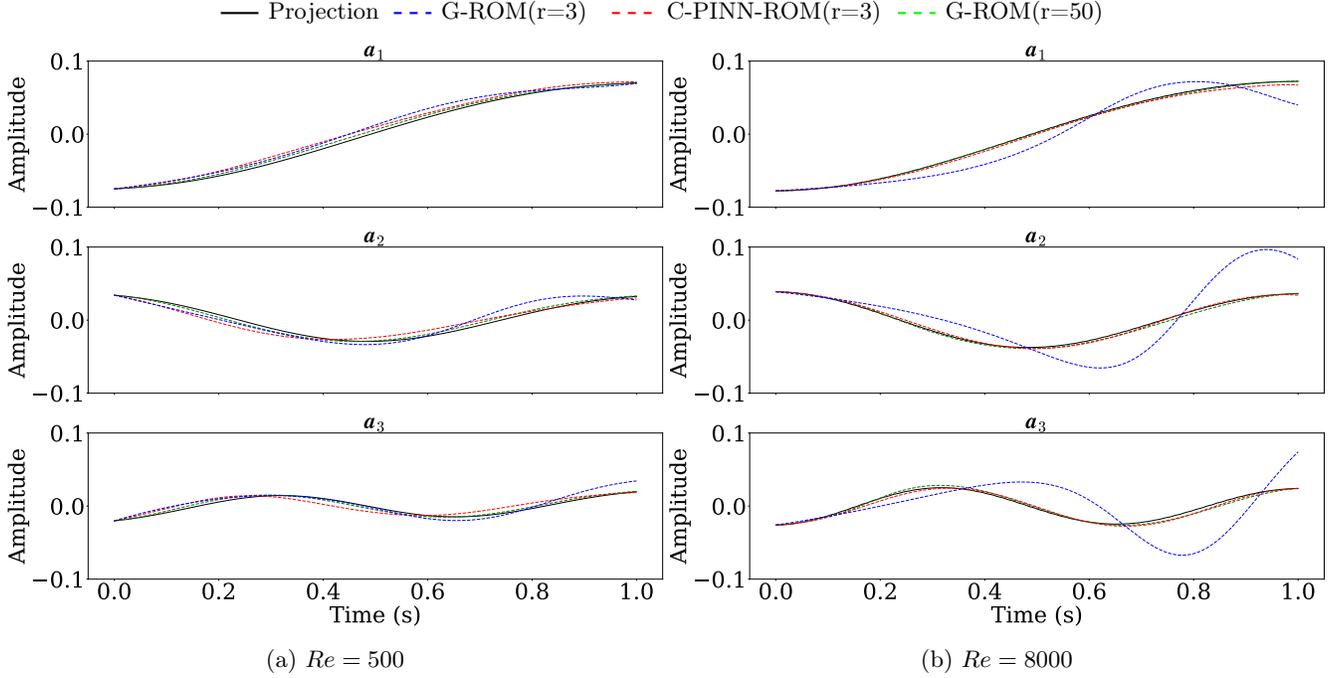
\begin{figure}[htbp!]
\centering
\begin{tikzpicture}
    \draw[thick, black] (0,0) -- (0.5,0) node[right, black] {\small Projection};
    \draw[thick, blue, dashed] (2.3,0) -- (2.8,0) node[right, black] {\small G-ROM(r=3)};
    \draw[thick, red, dashed] (5.2,0) -- (5.8,0) node[right, black] {\small C-PINN-ROM(r=3)};
    \draw[thick, green, dashed] (9.0,0) -- (9.5,0) node[right, black] {\small G-ROM(r=50)};
\end{tikzpicture}
\begin{subfigure}{.50\textwidth}
\centering
\includegraphics[width=1\linewidth]{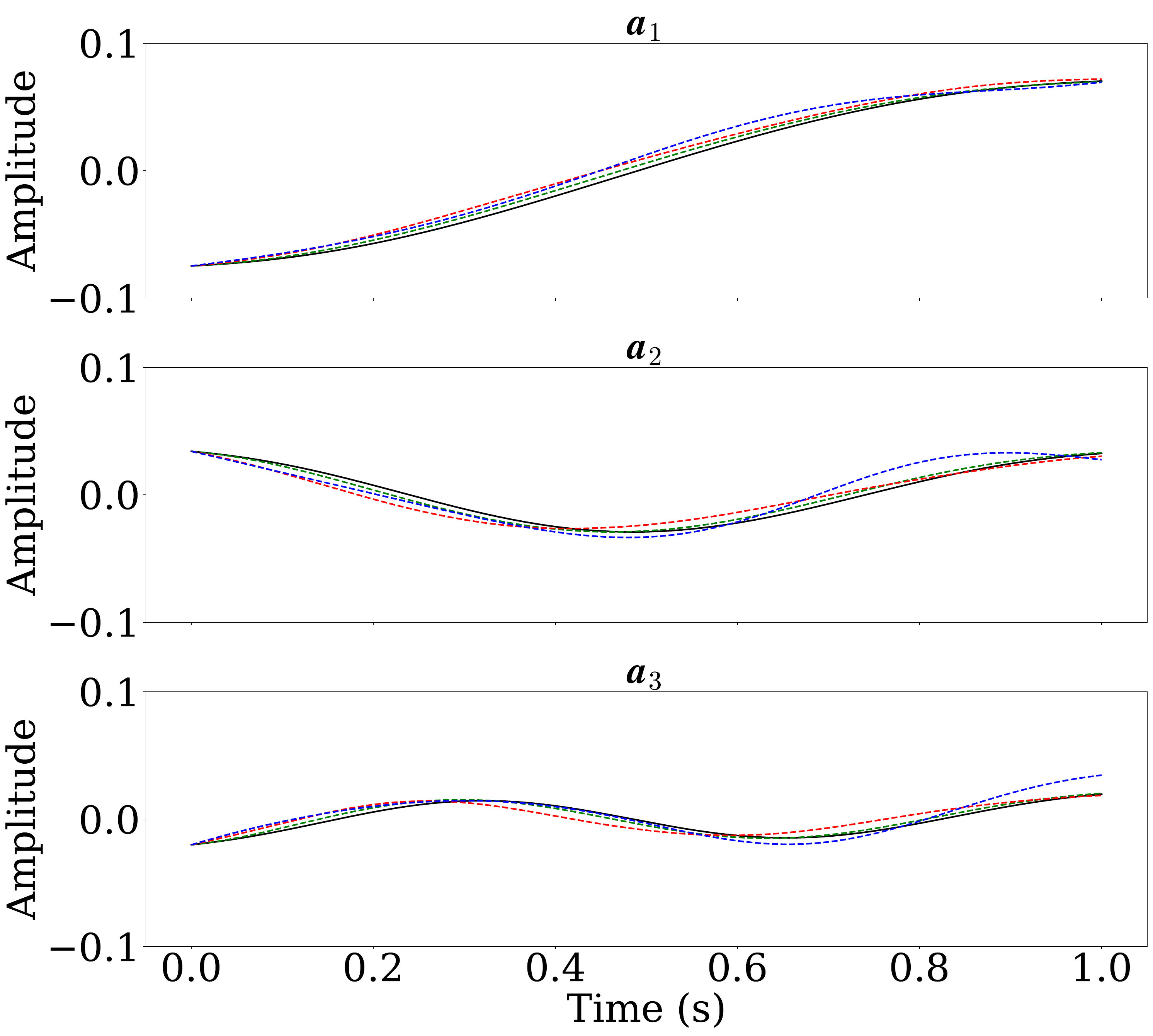}
\caption{$Re = 500$} \label{fig:10:500}
\end{subfigure}%
\begin{subfigure}{.50\textwidth}
\centering
\includegraphics[width=1\linewidth]{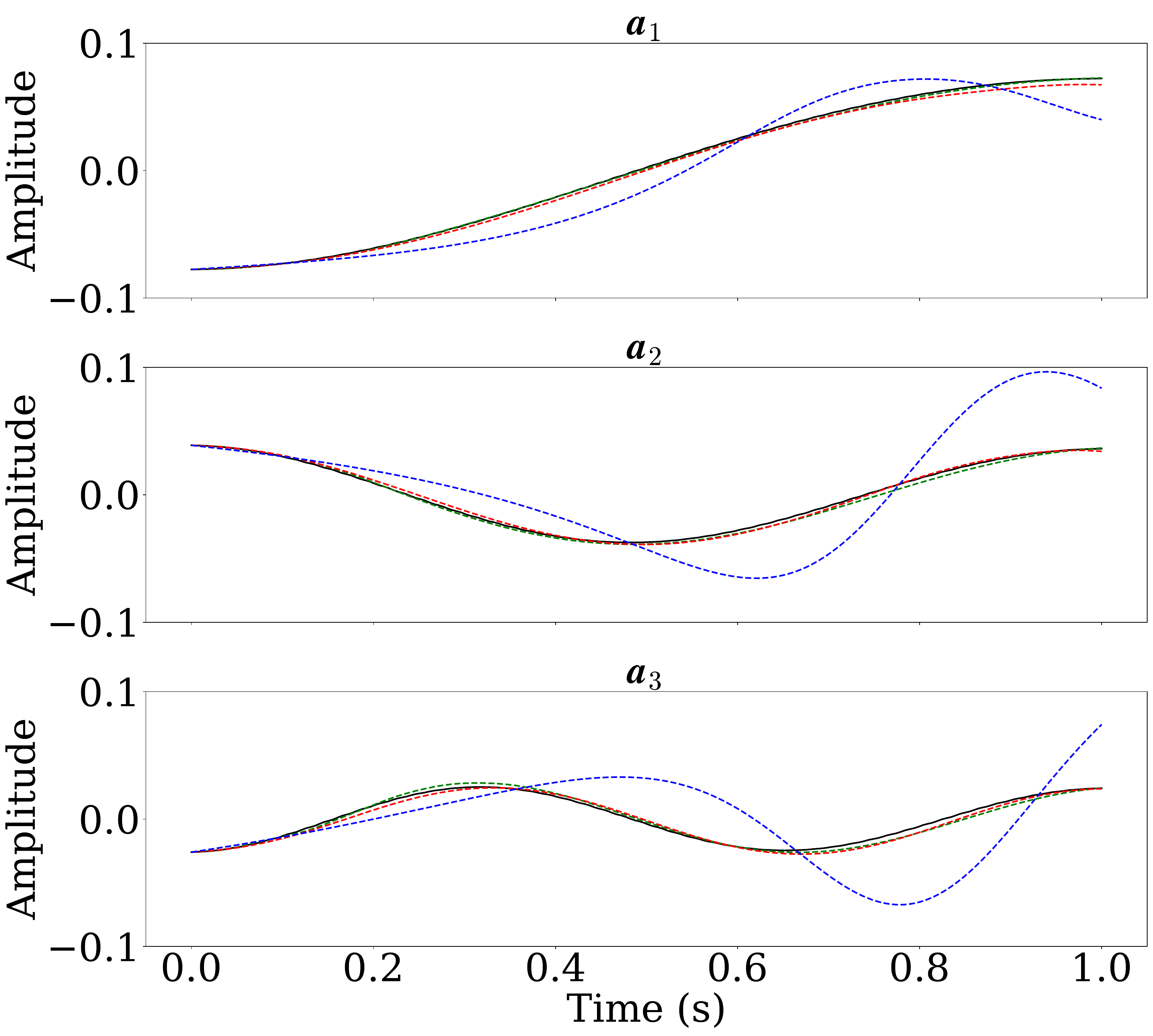}
\caption{$Re = 8000$} \label{fig:10:8000}
\end{subfigure}
\caption{Burgers equation; time-dependent coefficients in the extrapolation testing regime with $Re = 500$ and $Re=8000$. } \label{fig:Extrapolation}
\end{figure}

Figures~\ref{fig:Interpolation} and~\ref{fig:Extrapolation} illustrate the coefficient trajectories for the different ROM models. Examination of these trajectories shows that the C-PINN-ROM model provides a substantial improvement over the standard G-ROM with $r=3$ in both the interpolation and extrapolation testing regimes. Furthermore, the improvements of the C-PINN-ROM are even more pronounced when compared with higher-dimensional G-ROM simulations, in which additional truncated modes are resolved up to the model dimension $r$.

\begin{figure}[htbp!]
\centering
\begin{subfigure}{.48\textwidth}
\centering
\includegraphics[width=1\linewidth]{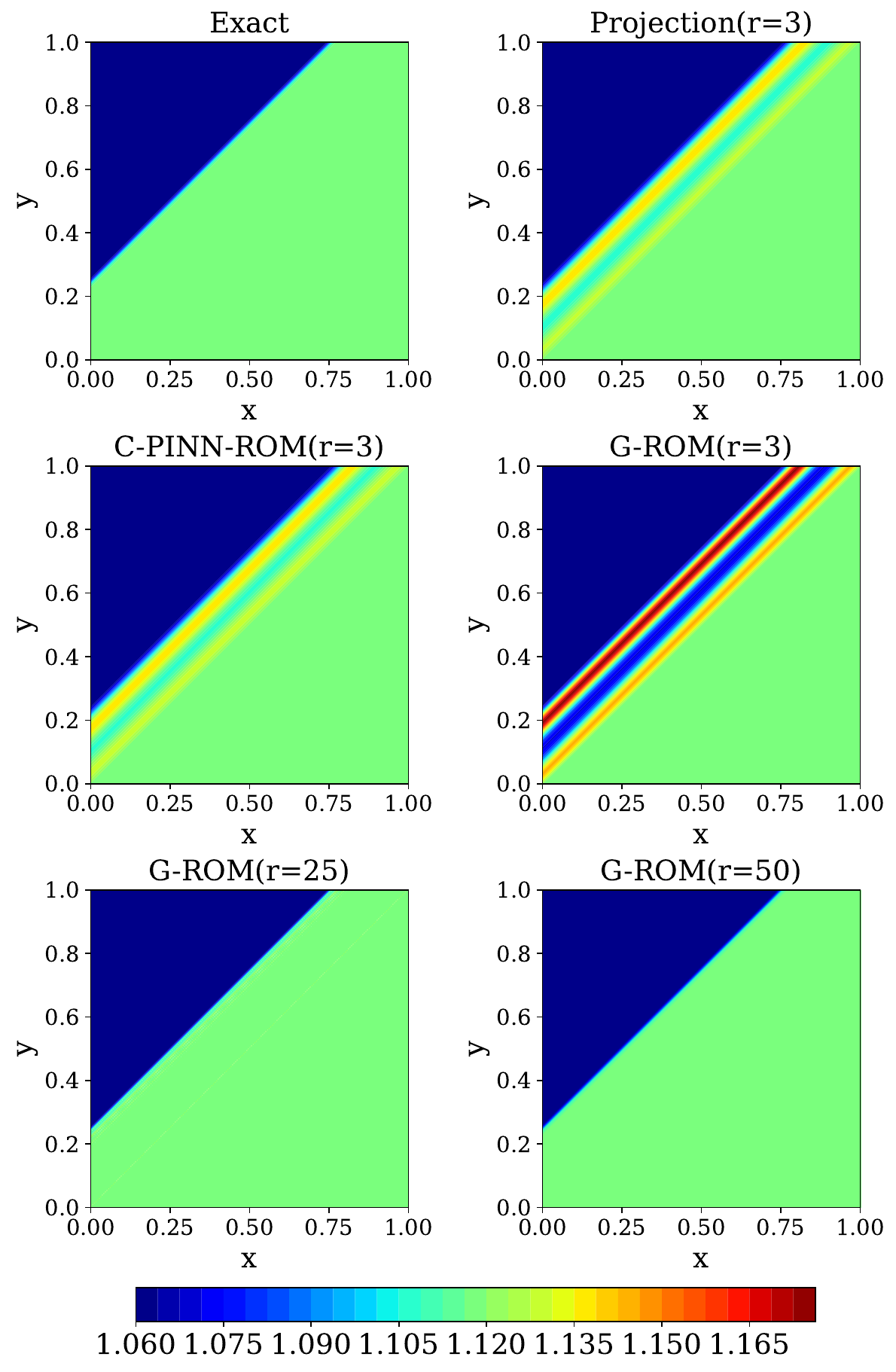}
\caption{$Re = 1500$} \label{fig:9:sub1}
\end{subfigure}%
\begin{subfigure}{.48\textwidth}
\centering
\includegraphics[width=1\linewidth]{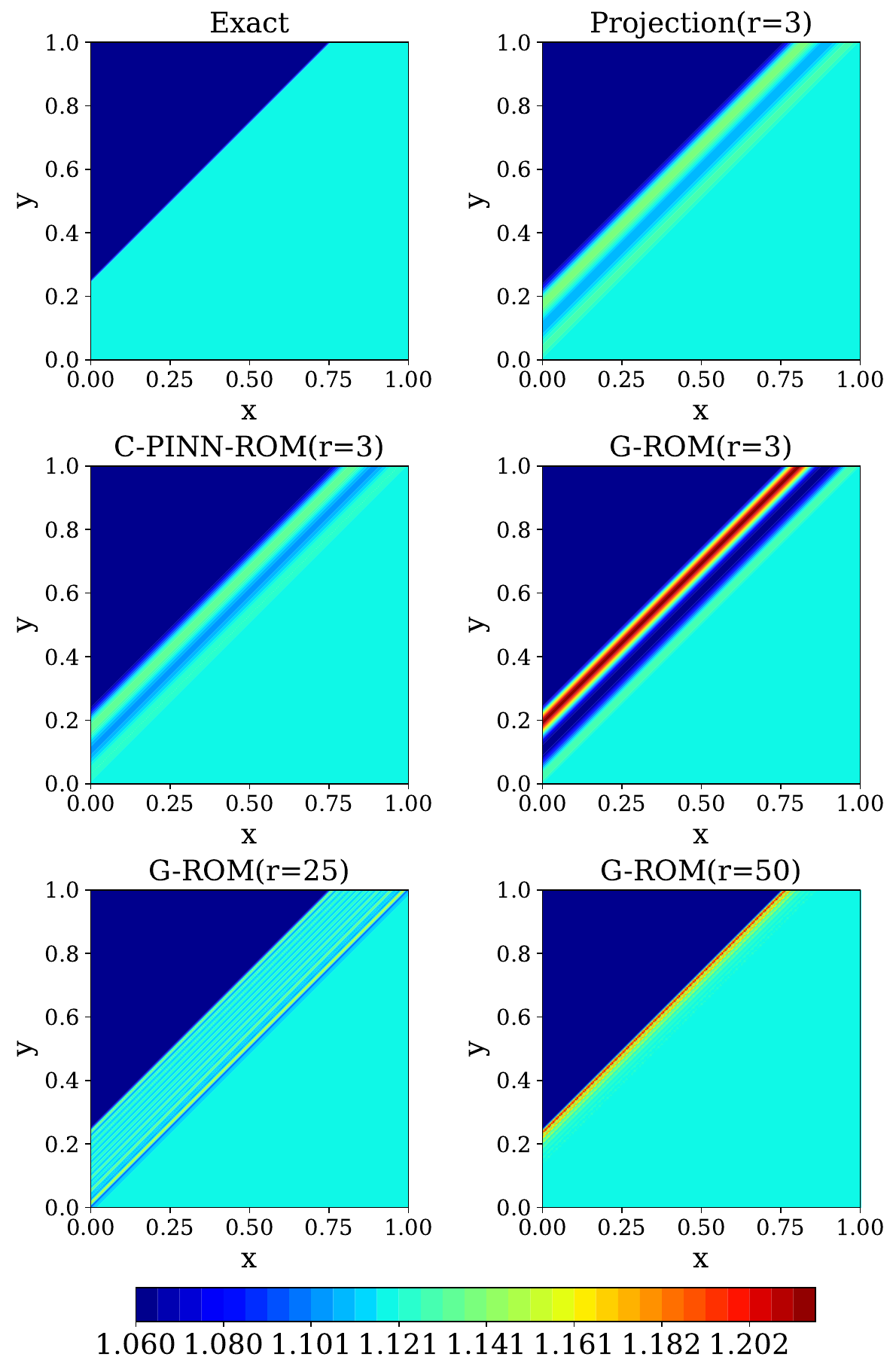}
\caption{$Re = 5500$} \label{fig:9:sub2}
\end{subfigure}
\caption{Burgers equation; solution field comparison in interpolation regime at final time step, with $Re = 1500$ shown on the left and $Re=5500$ on the right.} \label{fig:9}
\end{figure}

\begin{figure}[htbp!]
\centering
\begin{subfigure}{.48\textwidth}
\centering
\includegraphics[width=1\linewidth]{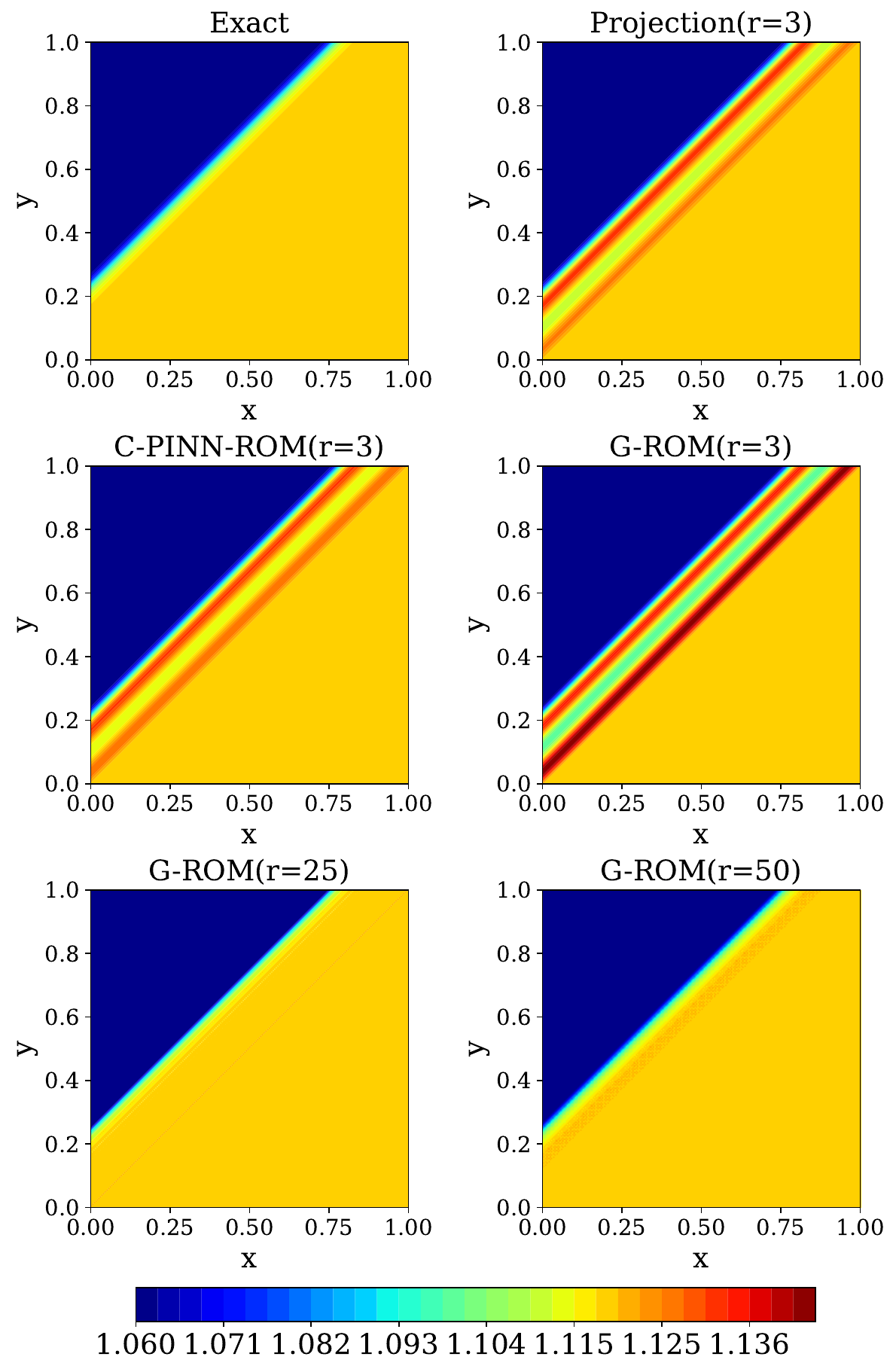} \caption{$Re = 500$} \label{fig:10:sub1}
\end{subfigure}%
\begin{subfigure}{.48\textwidth}
\centering
\includegraphics[width=1\linewidth]{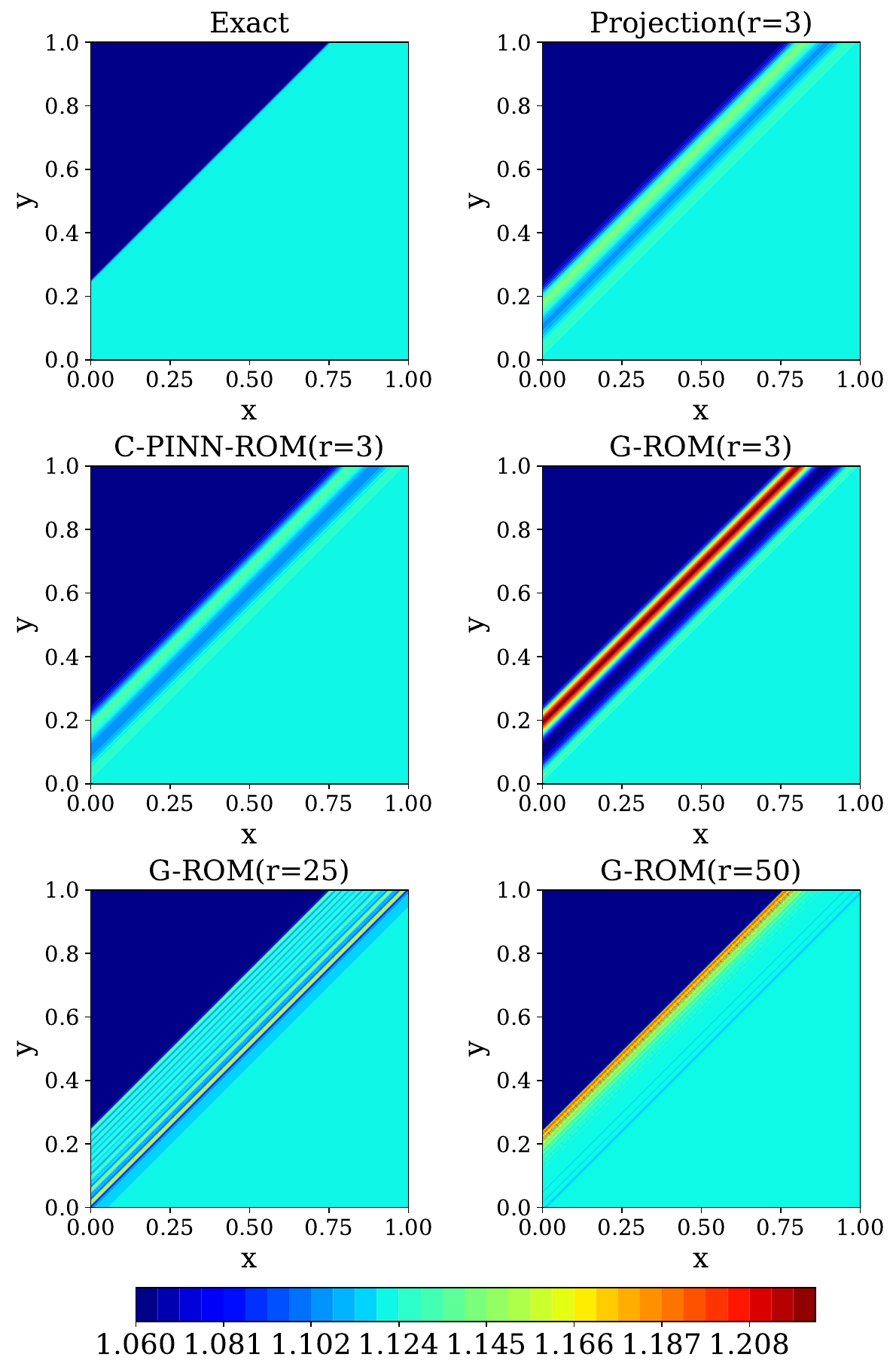}
\caption{$Re = 8000$} \label{fig:10:sub2}
\end{subfigure}
\caption{Burgers equation; solution field comparison in the extrapolation regime at final time step, with $Re = 500$ shown on the left and $Re=8000$ on the right.} \label{fig:10}
\end{figure}

Table~\ref{tab:rom_errors_final_time_updated} and Figures~\ref{fig:9} and~\ref{fig:10} present the relative $L^2$ errors and reconstructed solution fields at the final time for the interpolation and extrapolation regimes, respectively. This combined representation allows for both quantitative and qualitative assessment of the ROM models’ performance across different Reynolds numbers.

In the ROM space with $r=3$, the C-PINN-ROM significantly improves accuracy compared to the uncorrected G-ROM and closely matches the reference projection onto the same ROM space. The higher-dimensional G-ROMs with $r=25$ and $r=50$ achieve lower errors by capturing more energetic modes, but operate in larger ROM spaces and are not directly comparable to the $r=3$ projection. Notably, the C-PINN-ROM in the ROM space with $r=3$ attains accuracy approaching that of the $r=25$ G-ROM, highlighting the effectiveness of the closure model in accounting for the influence of truncated modes.

This trend is further confirmed by the solution fields at the final time. In the interpolation regime (Figure~\ref{fig:9}) with $Re=1500$ and $Re=5500$, the uncorrected G-ROM exhibits strong oscillations, whereas the C-PINN-ROM reproduces the reference projection closely. Similarly, in the extrapolation regime (Figure~\ref{fig:10}) with $Re=500$ and $Re=8000$, the uncorrected G-ROM shows pronounced oscillations near discontinuities, which are effectively mitigated by the closure model. Collectively, these results demonstrate that the C-PINN-ROM can achieve high accuracy in a low-dimensional ROM space, reducing computational cost while preserving solution fidelity.

\begin{table}[htbp!]
\centering
\small
\begin{tabular}{c|c|c|c|c|c}
\hline 
\textbf{Re} & \textbf{G-ROM(r=3)} & \textbf{C-PINN-ROM(r=3)} & \textbf{Projection(r=3)} & \textbf{G-ROM(r=25)} & \textbf{G-ROM(r=50)} \\
\hline
500  & $2.82\times 10^{-2}$ & $2.45\times 10^{-2}$ & $2.44\times 10^{-2}$ & $1.35\times 10^{-2}$ & $7.58\times 10^{-3}$ \\
1500 & $4.93\times 10^{-2}$ & $2.82\times 10^{-2}$ & $2.82\times 10^{-2}$ & $5.66\times 10^{-3}$ & $2.05\times 10^{-3}$ \\
3500 & $6.62\times 10^{-2}$ & $3.08\times 10^{-2}$ & $3.08\times 10^{-2}$ & $8.54\times 10^{-3}$ & $1.26\times 10^{-2}$ \\
5500 & $7.22\times 10^{-2}$ & $3.18\times 10^{-2}$ & $3.15\times 10^{-2}$ & $1.55\times 10^{-2}$ & $2.53\times 10^{-2}$ \\
8000 & $7.58\times 10^{-2}$ & $3.22\times 10^{-2}$ & $3.18\times 10^{-2}$ & $2.26\times 10^{-2}$ & $3.71\times 10^{-2}$ \\
\hline
\end{tabular}
\caption{Burgers equation; relative $L^2$ errors of ROM models for different Reynolds numbers at the final time step.}
\label{tab:rom_errors_final_time_updated}
\end{table}

\subsection{2-Dimensional Flow Past a Cylinder}
\label{2d_fpc_section}
\begin{figure}[htbp!]
\centering
\includegraphics[width=0.9\linewidth]{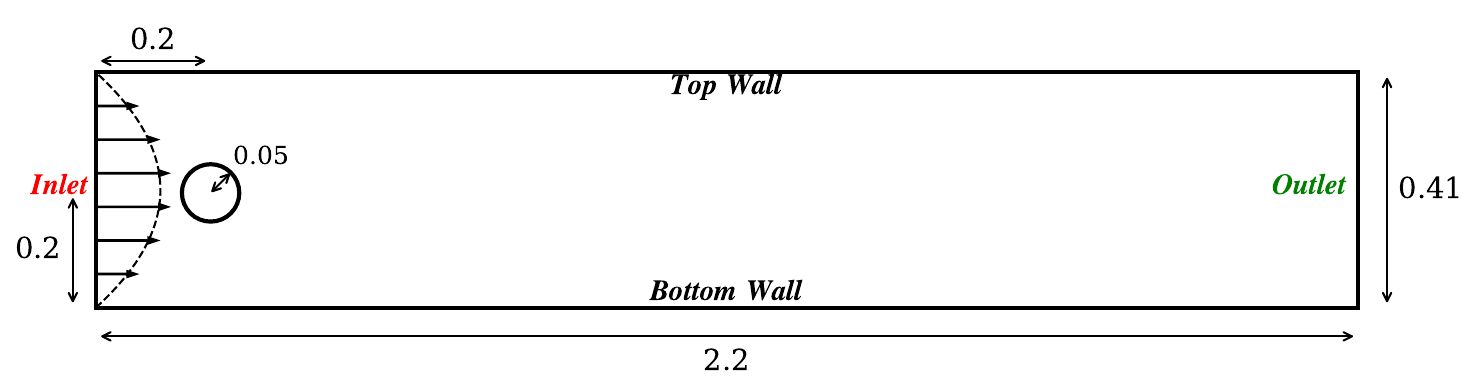}
\caption{Flow past a circular cylinder: computational domain used for the FOM simulations.} \label{fig:FPC_domain}
\end{figure}

In this section, the newly developed closure model, referred to as the C-PINN-ROM, is evaluated with respect to its temporal extrapolation capabilities. The benchmark problem considered is the incompressible flow past a circular cylinder at $Re = 1000$. The computational domain consists of a rectangular channel of dimensions $2.2 \times 0.41$, containing a circular cylinder of radius $r = 0.05$ located at $(0.2,\,0.2)$, as illustrated in Figure~\ref{fig:FPC_domain}.

The incompressible Navier--Stokes equations~\eqref{eqn:nse_momentum} and~\eqref{eqn:nse_incompressibility} are employed as the governing mathematical model. No-slip boundary conditions are imposed on the top and bottom channel walls, together with the inflow and outflow velocity boundary conditions given in~\eqref{eqn:FPC_BC_in} and~\eqref{eqn:FPC_BC_out}, respectively, following~\cite{john2004reference,mohebujjaman2017energy}.

\begin{align}
\label{eqn:FPC_BC_in}
u_1(0, y, t) &= u_1(2.2, y, t) = \frac{6}{0.41^2}\, y(0.41 - y), \\
\label{eqn:FPC_BC_out}
u_2(0, y, t) &= u_2(2.2, y, t) = 0 .
\end{align}


For the FOM simulations, the spatial domain is discretized using a barycenter-refined triangular mesh. On this mesh, the Scott--Vogelius finite element pair $(P_2, P^{\mathrm{disc}}_1)$ is employed, which ensures Ladyzhenskaya--Babuška--Brezzi (LBB) stability and enforces exact mass conservation through a pointwise divergence-free velocity field \cite{john2016divergence}. The computational mesh yields $98542$ velocity and $73341$ pressure degrees of freedom. The FOM simulations are performed with linearized BDF2 temporal discretization with $\Delta t = 0.002$, while for ROM simulation fourth-order Runge-Kutta scheme is utilized with the same $\Delta t=0.002$.

During the offline phase, FOM snapshots collected over the time interval $t \in [20,\,20.5]$ are used to construct the reduced-order models. The same snapshot data are employed both to generate the ROM basis and to train the closure term in the newly proposed C-PINN-ROM. For the flow past a cylinder benchmark problem, time is treated as the primary parameter of interest.

After constructing the ROM basis from these snapshots, the closure term is formulated using a variational multiscale (VMS) decomposition of the reduced space. Specifically, the resolved-scale subspace is chosen as $\dim(\boldsymbol{X}_r)=4$, while the unresolved-scale complementary subspace has dimension $\dim(\boldsymbol{X}_r^\perp)=30$.


The architectures of the neural networks used in the PINN-based closure model, along with the associated training hyperparameters, are summarized in Table~\ref{tab:pinn_fpc_arch}.

\begin{table}[htbp!]
\centering
\small
\begin{tabular}{l|c|c}
\hline
\textbf{Parameter} & \textbf{Network 1} & \textbf{Network 2} \\
\hline
Architecture & Fully Connected (MLP) & Fully Connected (MLP) \\
Layer Structure & $[2, 32, 32, 4]$ & $[4,128,128,128,128, 16]$ \\
Activation & $\tanh$ & $\tanh$ \\
Initialization & Xavier Normal &  Xavier Normal \\
\hline 
Initial Learning Rate & \multicolumn{2}{c}{$10^{-3}$} \\
Decay Rate / Steps & \multicolumn{2}{c}{$0.90$ / $5 \times 10^3$} \\
Optimizer & \multicolumn{2}{c}{Adam} \\
Training Steps & \multicolumn{2}{c}{$4\times10^5$} \\
\hline
\end{tabular}
\caption{Flow past a cylinder: network architectures and training hyperparameters of the PINN closure model $\mathbb{C}(\boldsymbol{a}_r)$.}
\label{tab:pinn_fpc_arch}
\end{table}

\begin{figure}[htbp!]
\centering
\includegraphics[width=0.60\linewidth]{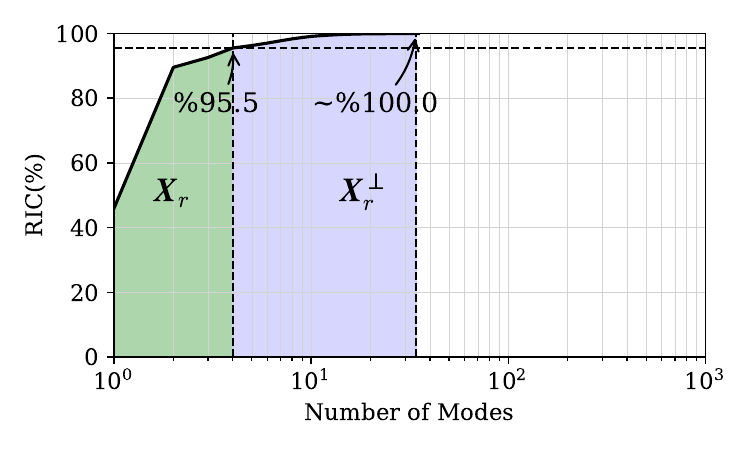}
\caption{Flow past a cylinder; relative information content of the complementary POD subspaces $\boldsymbol{X}_r$ and $\boldsymbol{X}_r^\perp$.} \label{fig:FPC_RIC}
\end{figure}

\begin{figure}[htbp!]
\centering
\includegraphics[width=0.75\linewidth]{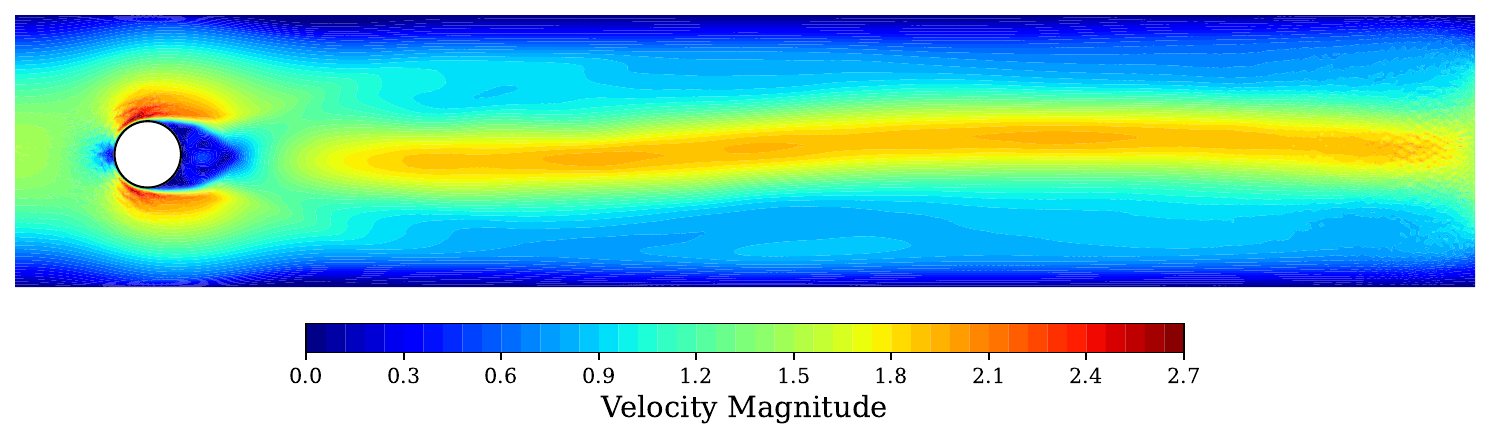}
\caption{Flow past a cylinder; mean velocity field of the snapshot data at $Re = 1000$.} \label{fig:FPC_meanfield}
\end{figure}

\begin{figure}[htbp!]
\centering
\includegraphics[width=1\linewidth]{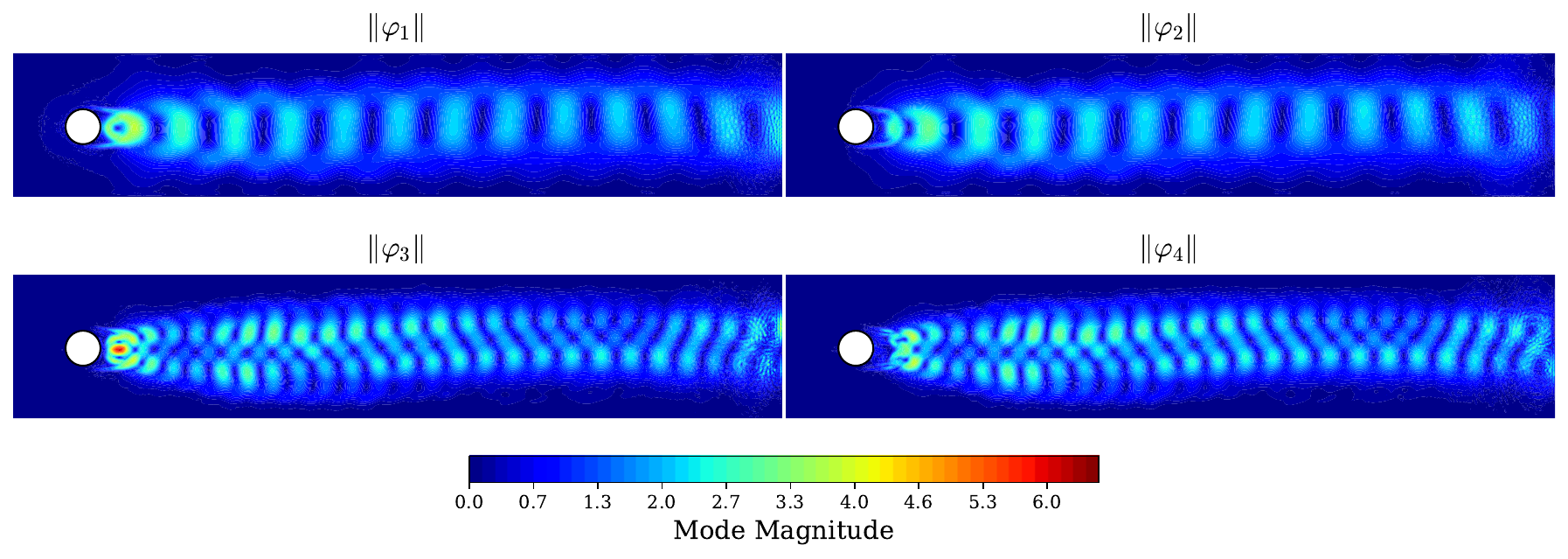}
\caption{Flow past a cylinder; velocity POD modes.} \label{fig:FPC_PODmodes}
\end{figure}

The energy distribution between the resolved and unresolved subspaces is shown in Figure~\ref{fig:FPC_RIC}. The resolved ROM space is fixed to $r=4$ in order to intentionally retain a significant portion of the flow energy in the unresolved subspace, which is subsequently modeled by the proposed closure term. The mean velocity field of the snapshot data and the POD modes spanning the velocity ROM
space are presented in Figures~\ref{fig:FPC_meanfield} and~\ref{fig:FPC_PODmodes},
respectively.

After the training phase, the newly proposed PINN closure model is evaluated over the time interval $t \in [20.5,\,23]$. For comparison, three different ROM simulations are performed.
First, the standard (uncorrected) G-ROM with $r=4$ is simulated to assess the effect of the closure model on time intervals not included in the training data. 
Next, G-ROM simulations with larger ROM spaces are conducted to examine the influence of the reduced-order subspace dimension on the predicted dynamics and overall solution accuracy.

\begin{figure}[htbp!]
\centering
\includegraphics[width=\linewidth]{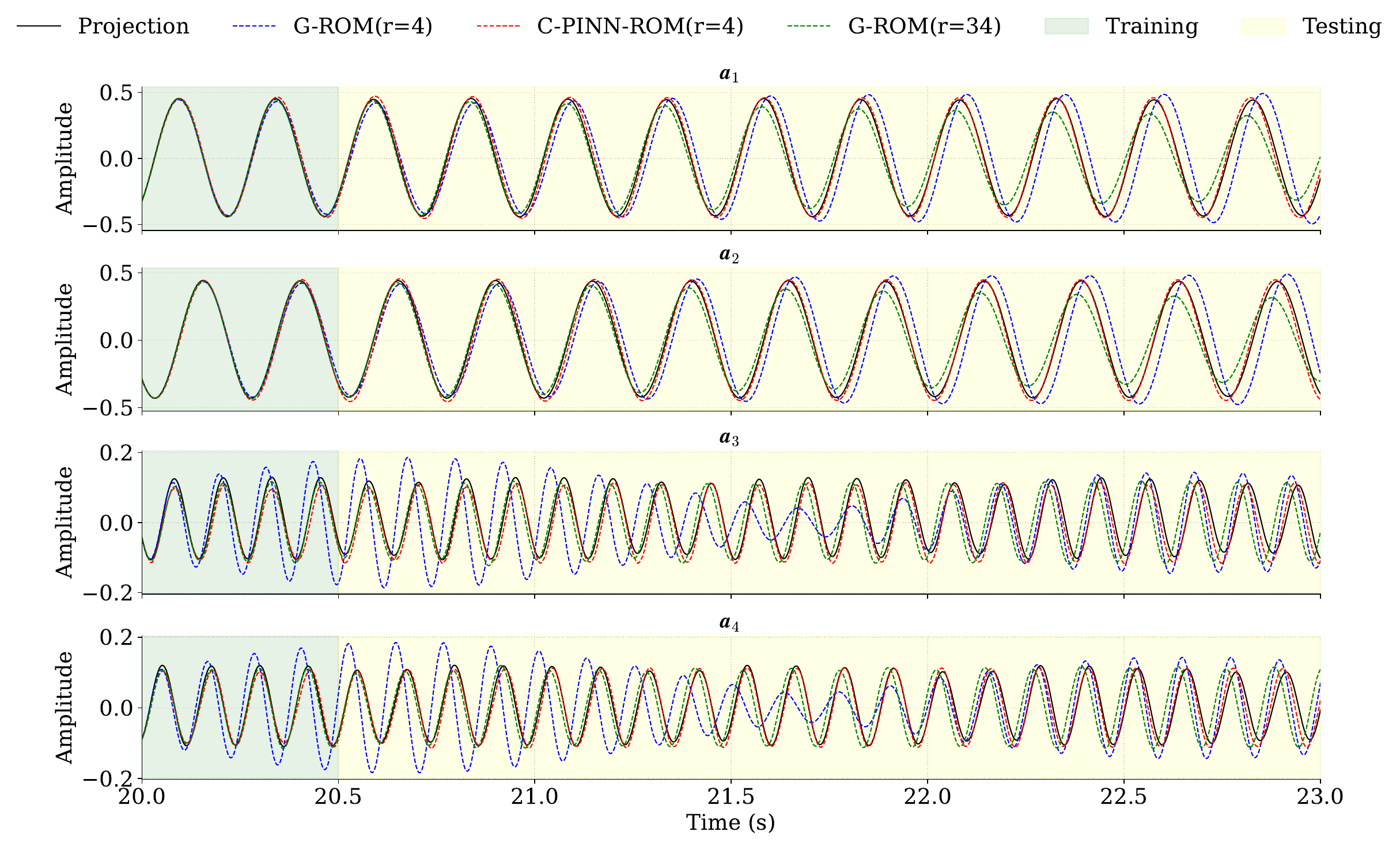}
\caption{Flow past a cylinder; time-dependent coefficient comparison over both the training and testing time intervals.} \label{fig:FPC_time_coeffs}
\end{figure}

Figure~\ref{fig:FPC_time_coeffs} compares the time evolution of the resolved ROM coefficients in the temporal extrapolation regime. The uncorrected G-ROM with $r=4$ exhibits noticeable deviations from the true coefficient trajectories for all modes, with discrepancies appearing even within the time interval used to construct the ROM
basis. This behavior highlights the limited predictive capability of the standard Galerkin ROM when applied outside its effective dynamical range.

%
\newpage
Further analysis of the G-ROM with $r=34$ reveals that, over time, the modal amplitudes drift from their true values, with a noticeable phase shift appearing in the third and fourth mode coefficients. In contrast, the G-ROM with $r=4$ augmented with the PINN closure model exhibits substantially improved accuracy across all modal trajectories. Remarkably, the closure-enhanced ROM produces more accurate trajectories than the higher-dimensional G-ROM ($r=34$), highlighting that simply increasing the ROM dimension does not guarantee improved time-coefficient fidelity.


In Table~\ref{tab:rom_ordered_comparison}, the relative $L^2$ errors of ROM models are listed for varied ROM dimensions $r = [4, 8, 12, 16, 20, 24, 28, 32, 34]$, including the C-PINN-ROM ($r=4$) and the reference projection onto the resolved subspace. Both mean and final-time errors are presented to evaluate the accuracy of the models over
the entire simulation interval and at the final time step. 

The results indicate that the C-PINN-ROM applied to the uncorrected G-ROM ($r=4$) achieves the lowest errors in both metrics, closely matching the reference projection. Notably, simply increasing the ROM dimension in the standard G-ROM does not consistently reduce the relative $L^2$ errors, highlighting the effectiveness of the closure model in capturing the influence of truncated modes and improving the fidelity of the reduced-order trajectories.

\begin{table}[htbp!]
\centering
\small
\begin{tabular}{c|c|c}
\hline 
\textbf{Model} & \textbf{Mean} & \textbf{Final Time} \\
\hline
G-ROM(r=4)  & $2.20\times 10^{-1}$ & $2.95\times 10^{-1}$ \\
C-PINN-ROM(r=4)   & $1.14\times 10^{-1}$ & $1.30\times 10^{-1}$ \\
Projection(r=4)  & $1.08\times 10^{-1}$ & $1.08\times 10^{-1}$ \\
G-ROM(r=8)  & $1.42\times 10^{-1}$ & $2.06\times 10^{-1}$ \\
G-ROM(r=12) & $2.05\times 10^{-1}$ & $3.67\times 10^{-1}$ \\
G-ROM(r=16) & $1.80\times 10^{-1}$ & $2.55\times 10^{-1}$ \\
G-ROM(r=20) & $1.80\times 10^{-1}$ & $2.55\times 10^{-1}$ \\
G-ROM(r=24) & $1.48\times 10^{-1}$ & $2.11\times 10^{-1}$ \\
G-ROM(r=28) & $1.58\times 10^{-1}$ & $2.96\times 10^{-1}$ \\
G-ROM(r=32) & $1.49\times 10^{-1}$ & $2.27\times 10^{-1}$ \\
G-ROM(r=34) & $1.43\times 10^{-1}$ & $2.66\times 10^{-1}$ \\
\hline
\end{tabular}
\caption{Flow past a cylinder; relative $L^2$ errors of ROM models for varied ROM dimension $r$, presented in the order of experiments.}
\label{tab:rom_ordered_comparison}
\end{table}

\begin{figure}[htbp!]
\centering
\includegraphics[width=1\linewidth]{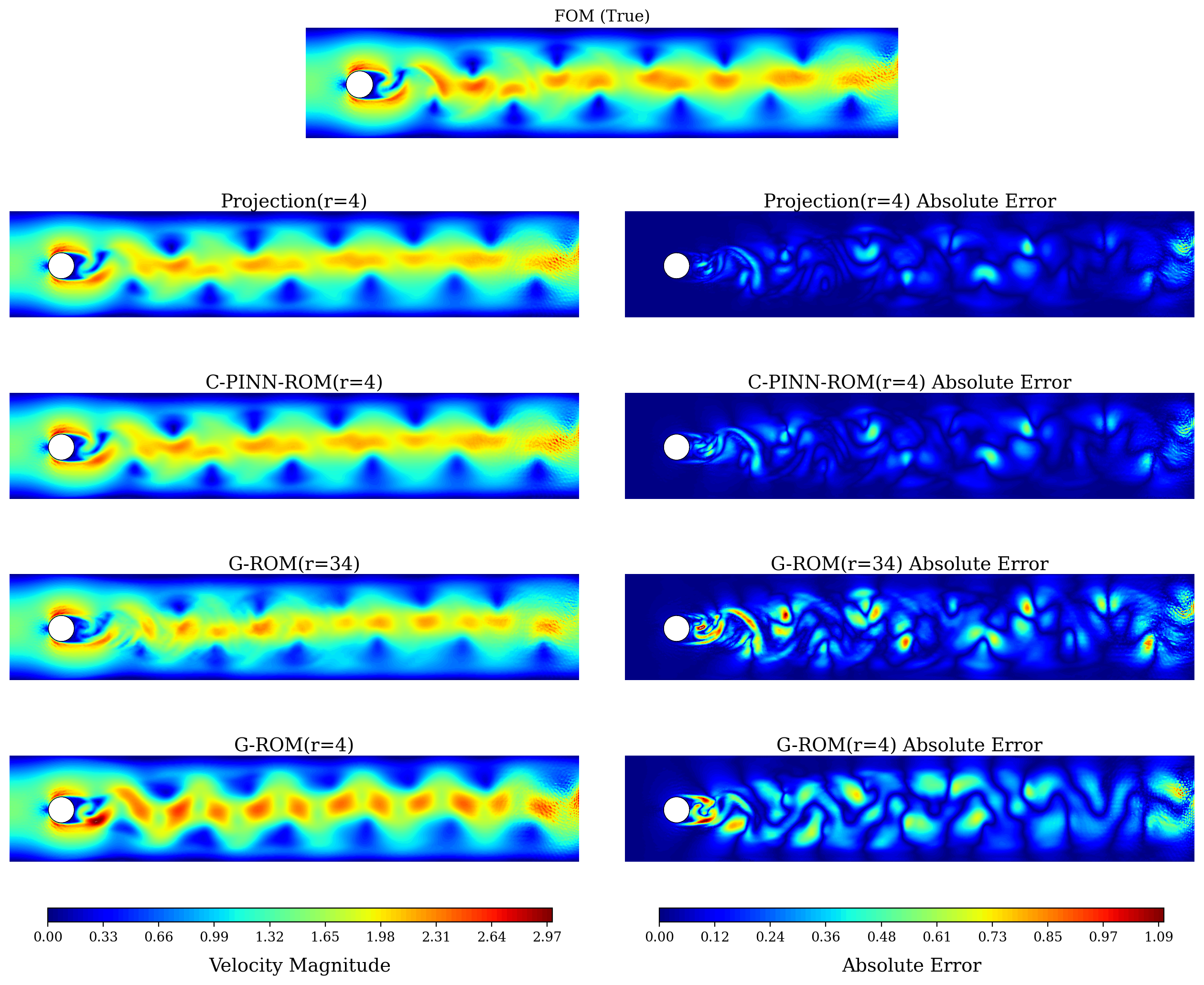}
\caption{Flow past a cylinder; solution and absolute error fields at the final time step $t = 23$ s for different models.} \label{fig:FPC_solutions}
\end{figure}

%
Figure~\ref{fig:FPC_solutions} presents the reconstructed velocity fields of the different ROM models alongside their corresponding relative $L^2$ error fields (as defined in \eqref{eq:rel_l2_fom_err}) at the final time step $t = 23$ s. While the projection of the FOM solution onto the resolved subspace provides the most accurate representation, it is evident that the C-PINN-ROM ($r=4$) outperforms
both the uncorrected G-ROM and higher-dimensional Galerkin ROMs. The closure model effectively reduces the discrepancy with the FOM solution, capturing the dynamics missed by the truncated modes in the standard ROM simulations.

\section{Conclusions and Outlook} 
\label{sec:conclusions_outlook}

In this work, a PINN closure model is developed to enhance under-resolved Galerkin ROM simulations of convection-dominated flows. The newly proposed closure model is derived from a VMS formulation of the reduced dynamical system. By treating the exact closure terms as reference data, the ROM ODE system augmented with the predicted closure is incorporated directly as a constraint in the network optimization. Unlike existing data-driven ROM closure approaches, the key novelty of this study lies in explicitly embedding the governing equations into the training process of the closure model.


The performance of the proposed C-PINN-ROM framework is evaluated on two benchmark problems: the two-dimensional viscous Burgers equation and the two-dimensional flow past a circular cylinder. For the Burgers equation, the closure model is constructed in a parametric setting and tested under both interpolation and extrapolation of the Reynolds number. For the flow past a circular cylinder, the emphasis is placed on assessing the temporal extrapolation capabilities of the closure model.


In the parametric Burgers equation study, the PINN closure model significantly improves the accuracy of the standard Galerkin ROM within the same ROM space for both interpolation and extrapolation regimes, as quantified by deviations from the projected reference solution. Moreover, the C-PINN-ROM achieves accuracy comparable to that of higher-dimensional Galerkin ROMs in terms of relative $L^2$ errors, i.e., $\mathcal{E}(L^2)$, demonstrating that the proposed closure model can deliver similar accuracy at a substantially lower computational cost.


For the flow past a circular cylinder, the closure model effectively corrects the G-ROM when the reduced-order trajectories deviate from the reference solution during temporal extrapolation. Importantly, increasing the dimension of the ROM subspace alone does not lead to improved accuracy compared to either the projected solution or the C-PINN-ROM results, underscoring the efficiency and robustness of the closure-based approach.


In the present study, the closure model’s parametric and temporal extrapolation capabilities were examined separately across different benchmark problems. As a direction for future work, emphasis will be placed on extending the proposed framework to simultaneously address both parametric and temporal extrapolation within a single problem setting. In addition, further research will focus on the treatment of projection errors in G-ROM systems. Since projection errors are intrinsic to the reduced subspace and independent of the reduced dynamical system, they impose a fundamental limitation on ROM accuracy. Future efforts will therefore explore physics-based correction strategies aimed at mitigating projection-induced errors and further improving the predictive capability of reduced-order models.

\paragraph{Acknowledgments:} 
The second author is partially supported by Project PID2021-123153OB-C21 funded by MCIN/AEI/10.13039/501100011033/FEDER, UE.

\section*{Statements and Declarations}

\subsection*{Competing Interests}
We wish to confirm that there are no known conflicts of interest associated with this publication and that there has been no significant financial support for this work that could have influenced its outcome. Furthermore, the authors have no relevant financial or nonfinancial interests to disclose.

\subsection*{Data Availability}
Data will be made available on request. 

\subsection*{Author Contributions}
We confirm that the manuscript has been read and approved by all named authors and that there are no other persons who meet the authorship criteria but are not listed. We further confirm that all have approved the order of authors listed in the manuscript of us. 

We confirm that we have given due consideration to the protection of intellectual property associated with this work and that there are no impediments to publication, including the timing of publication, with respect to intellectual property. In so doing, we confirm that we have followed the regulations of our institutions concerning intellectual property.

We understand that the Corresponding Author is the sole contact for the Editorial process (including Editorial Manager and direct communications with the office). The corresponding author is responsible for communicating with the other authors about progress, submissions of revisions, and final approval of proofs. We confirm that we have provided a current, correct email address accessible by the Corresponding Author. 
\newpage
\bibliographystyle{abbrv}
\bibliography{reference}

\end{document}